\documentclass[a4paper,11pt,pdftex]{article}
\pdfoutput=1

\usepackage{slashed}
\usepackage{braket}
\usepackage{jheppub}
\usepackage{graphicx}
\usepackage{url}
\usepackage{amssymb,amsmath}
\usepackage{float}
\usepackage{wrapfig}
\usepackage{bm}
\usepackage{CJKutf8}
\usepackage{color}
\usepackage{comment}
\usepackage{xcolor}
\usepackage[normalem]{ulem}
\usepackage{hyperref}
\usepackage{mathrsfs}
\usepackage[version=4]{mhchem}
\usepackage{booktabs}
\usepackage[noabbrev]{cleveref}
\hypersetup{
colorlinks=true,
citecolor=blue,
citebordercolor=red,
linkcolor=blue,
urlcolor=blue
}

\catcode`\@=11
\catcode`\@=12

\newcommand{\p}{\partial}

\newcommand{\bp}{\begin{pmatrix}}
\newcommand{\ep}{\end{pmatrix}}

\newcommand{\bs}[1]{\boldsymbol}

\newcommand{\be}{\begin{equation}}
\newcommand{\ee}{\end{equation}}

\newcommand{\ba}{\begin{array}} 
\newcommand{\ea}{\end{array}}

\graphicspath{{./figs/}}
\newbox{\ORCIDicon}

\makeatletter
\gdef\@fpheader{\phantom{prepared for submission to JHEP}}
\makeatother

\makeatletter
\@addtoreset{equation}{section}
\makeatother

\begin{document}
\begin{flushright} 
YGHP-24-02
\end{flushright} 
\title{
Topological solitons stabilized by a background gauge field and
soliton-anti-soliton asymmetry 
}
\author[a,b]{Yuki Amari,}
\emailAdd{amari.yuki@keio.jp}

\author[c,a,d]{Minoru Eto,}
\emailAdd{meto@sci.kj.yamagata-u.ac.jp}

\author[b,a,d]{Muneto Nitta}
\emailAdd{nitta@phys-h.keio.ac.jp}

\affiliation[a]{Research and Education Center for Natural Sciences, Keio University, 4-1-1 Hiyoshi, Yokohama, Kanagawa 223-8521, Japan}

\affiliation[b]{Department of Physics, Keio University, 4-1-1 Hiyoshi, Yokohama, Kanagawa 223-8521, Japan}

\affiliation[c]{Department of Physics, Yamagata University, Kojirakawa-machi 1-4-12, Yamagata, Yamagata 990-8560, Japan}

\affiliation[d]{
International Institute for Sustainability with Knotted Chiral Meta Matter(SKCM$^2$), Hiroshima University, 1-3-2 Kagamiyama, Higashi-Hiroshima, Hiroshima 739-8511, Japan
}

\abstract{
We study topological lumps supported by 
the second homotopy group $\pi_2(S^2) \simeq {\mathbb Z}$ in 
a gauged $O(3)$ model without any potential term 
coupled with a (non)dynamical $U(1)$ gauge field. 
It is known that 
gauged-lumps are stable with an easy-plane potential term 
but are unstable to expand if the model has no potential term.
In this paper, we find that 
these gauged lumps without a potential term can be made stable 
by putting them in a uniform magnetic field, 
irrespective of whether the gauge field is dynamical or not. 
In the case of the non-dynamical gauge field,  only either of lumps or anti-lumps stably exists depending on the sign of the background magnetic field, and the other is unstable to shrink to be singular. 
We also construct coaxial multiple lumps 
whose size and mass exhibit a behaviour of droplets. 
In the case of the dynamical gauge field, 
both the lumps and anti-lumps 
stably exist with different masses; 
the lighter (heavier) one corresponds to 
the (un)stable one in the case of the nondynamical gauge field.
We find that a lump behaves as a superconducting ring 
and traps magnetic field in its inside, 
with the total magnetic field reduced from 
the background magnetic field.
}


\maketitle 

\section{Introduction}

Solitons are non-perturbative excitations playing significant roles 
in physics and mathematics.
In particular, topological solitons are protected by topology and thus are quite stable, ubiquitously appearing  
in quantum field theory \cite{Rajaraman:1987,Manton:2004tk,Shnir:2005vvi,Eto:2006pg,Vachaspati:2006zz,Weinberg:2006rq,
Shifman:2009zz,
Dunajski:2010zz,Weinberg:2012pjx,Shnir:2018yzp,Eto:2013hoa}, 
cosmology \cite{Kibble:1976sj,Vilenkin:1984ib,Hindmarsh:1994re,Vilenkin:2000jqa,Vachaspati:2015cma} 
and condensed matter systems 
\cite{Mermin:1979zz,Volovik:2003fe,Pismen,Svistunov:2015,Kawaguchi:2012ii}.
Topological solitons are classified into defects and textures, 
the both of which are characterized by homotopy groups 
in a different fashion.
The former contains domain walls, vortices and monopoles, while the latter does 
sine-Gordon solitons, baby(2D) Skyrmions, and Skyrmions. 
However, topology is not sufficient for the stability of solitons; 
energetically they may prefer to shrink or expand without suitable settings.
The Derrick's scaling argument offers a simple criterion of 
the stability under scaling of solitons of finite energy \cite{Derrick:1964ww}. 
For instance, in $d =3+1$ dimensions, 
Skyrmions \cite{Skyrme:1961vq,Skyrme:1962vh} and Hopfions \cite{Gladikowski:1996mb,Faddeev:1996zj,Battye:1998pe} 
are unstable to shrink without 
a four derivative Skyrme term.
In $d= 2+1$ dimensions, topological lumps (equivalently 
sigma model instantons in 2+0 dimensions) 
are scale invariant \cite{Polyakov:1975yp}, 
while they shrink in the presence of a potential term,  
which can be prevented by introducing 
a four-derivative Skyrme term 
as baby Skyrmions \cite{Bogolubskaya:1989ha,Bogolyubskaya:1989fz,Piette:1994ug,Weidig:1998ii} 
or time dependence
as Q-lumps \cite{Leese:1991hr,Abraham:1991ki}.
Vortices can be made finite energy by introducing gauge field,
otherwise they have infinite energy.

Here, we pursue a yet another possibility 
for the stabilization of topological solitons
in terms of a background field. 
Topological solitons unstable to shrink or expand 
can be prevented from collapsing by a background field. 
In fact, 
such a stabilization mechanism is well known in condensed matter physics
as far as background fields are not dynamical, 
while 
our target is the case of dynamical fields with backgrounds; 
In condensed matter physics,  
a typical example with nondynamical background field 
is given by magnetic Skyrmions 
\cite{Bogdanov:1989,Bogdanov:1995,Rossler:2006,doi:10.1126/science.1166767,doi:10.1038/nature09124,Han:2010by,
Barton-Singer:2018dlh,Ross:2020hsw,Amari:2023gqv}, that is 
2D Skyrmions in chiral magnets, which are stable even 
without a four derivative Skyrme term. 
Chiral magnets are
described by the $\mathbb{C}P^1$ model 
with a so-called Dzyaloshinskii-Moriya (DM) interaction \cite{Dzyaloshinskii,Moriya:1960zz} 
which prevents Skyrmions from shrinking.
This interaction can be reformulated as a background $SU(2)$ gauge field 
(a constant $SU(2)$ magnetic field) \cite{Schroers:2019hhe,Amari:2023gqv}.  
See e.g.~Refs.~\cite{Ross:2022vsa,Amari:2023gqv} for 
a Derrick's scaling argument. 
One of characteristic features is 
that only Skyrmions are stable 
while anti-Skyrmions are unstable to shrink.
Magnetic Skyrmions have been generalized to 
${\mathbb C}P^2$ Skyrmions \cite{Akagi:2021dpk,Amari:2022boe}.
Ultracold atomic gases with 
spin-orbit couplings 
or synthetic gauge fields \cite{Lin:2009vfu,Goldman:2013xka}
offer various examples 
such as 
2D Skyrmions \cite{PhysRevA.84.011607} 
and
3D Skyrmions
\cite{Kawakami:2012zw}.
As mentioned above, 
in all condensed matter examples, a gauge field is nondynamical 
and thus is a constant background field.

Let us illustrate our idea with taking  
a concrete example of topological lumps 
(or sigma model instantons in $d=2+0$) supported by 
the second homotopy group $\pi_2(S^2) \simeq {\mathbb Z}$ in 
an $O(3)$ model or equivalently ${\mathbb C}P^1$ model 
in $d=2+1$ dimensions 
\cite{Polyakov:1975yp}.
Topological lumps are scale invariant solutions in the ${\mathbb C}P^1$ model without a potential.
The structure of topological lumps becomes completely different if we gauge an $SO(2) \simeq U(1)$ subgroup of $O(3)$.
The $U(1)$ gauged ${\mathbb C}P^1$ model with an easy-plane potential term 
is known to admit stable lumps 
\cite{Schroers:1995he,Schroers:1996zy,Baptista:2004rk,Nitta:2011um}, 
where a lump is decomposed into 
a pair of a vortex and anti-vortex with fractional lump charges 
(sometimes called merons).
However, if there is no potential term, 
lumps are unstable to expand and eventually diluted.

In this paper, 
 we show that 
these gauged-lumps without potentials can be made stable 
if we put them in a uniform magnetic field background.
We separately consider the cases that the $U(1)$ gauge field is nondynamical and dynamical, 
and construct topological lump solutions in the presence of a uniform magnetic field background. 
In the both cases,
we apply the Derrick's scaling argument,  
numerically solve the equations of motion (EOM) for the axisymmetric lumps, 
and confirm the Derrick's scaling argument 
within numerical accuracy. 
One of interesting results peculiar to our models with the background magnetic field is asymmetry between solitons and anti-solitons;  
In the case of the non-dynamical gauge field,  either of lumps or anti-lumps stably exists depending on the sign of the background magnetic field, and the other is unstable to shrink to a singular configuration. 
This situation is quite analogous to 
magnetic Skyrmions in chiral magnets.
We find that the stable (anti-)lumps have no size modulus reflecting the fact that the model is not scale invariant due to the mass scale $\sqrt{eB}$ with the background magnetic field $B$  and  $U(1)$ gauge coupling constant $e$. 
We numerically find that the lump size is of order $1/\sqrt{eB}$ and asymptotic tails of profile functions decay to the vacuum 
exponentially, unlike ungauged BPS lumps 
whose profiles decay polynomially. 
The mass of the single lump is smaller than that of a BPS lump. 
We also construct coaxial lumps with higher winding number $k$ 
and find that their size and mass are proportional to $\sqrt{k}$ up to a constant shift, implying that the lumps behave as droplets. 
The higher lumps exhibit ring-like shape and their interiors are empty. 
On the other hand,
in the case of the dynamical gauge field, 
the above properties of the stable lumps for the non-dynamical gauge field remain qualitatively correct.  
However, unlike the nondynamical case, both the lumps and anti-lumps 
stably exist as regular solutions with different masses; 
the lighter one corresponds to the regular one 
in the case of the nondynamical gauge field, 
while the heavier one does to the unstable (singular) one.
The dynamical gauge field deforms the magnetic field around the lumps. 
We find that a lump behaves as a superconducting ring 
and traps magnetic field in its inside, 
with the total magnetic field reduced from 
the background magnetic field because of
superconductiviy expelling it.
The higher lumps appear as ring-like shape but their interiors are not empty and are filled by trapped magnetic fields which are almost constant.

A motivation to consider the problem of this paper is originated from QCD 
in the presence of strong magnetic fields.
QCD phase structure with a finite chemical potential 
in the presence of strong magnetic fields 
gather a lot of attention due to 
its relevance in neutron stars and heavy-ion collisions  
\cite{Kharzeev:2015kna,Miransky:2015ava,Andersen:2014xxa,Yamamoto:2021oys,Cao:2021rwx,Iwasaki:2021nrz}.
When a magnetic field 
and/or chemical potential is large enough, 
a neutral pion $\pi^0$ domain wall or soliton
has a negative tension due to the anomalous term \cite{Son:2007ny} 
when it is perpendicular to the magnetic field \cite{Eto:2022lhu}.
Thus, domain walls are spontaneously created in such a region, 
and the ground state is a stack of the domain walls, 
called a chiral soliton lattice (CSL)~\cite{Eto:2012qd,Brauner:2016pko}.
However, it has been recently found that 
a large part of CSL with stronger magnetic field and larger baryon density
should be replaced by a new phase, 
the domain-wall Skyrmion phase \cite{Eto:2023lyo,Eto:2023wul} (see also Ref.~\cite{Eto:2023tuu} for a counterpart under rapid rotation),
where topological lumps 
supported by $\pi_2(S^2) \simeq {\mathbb Z}$
have negative energy and spontaneously appear inside a soliton 
\cite{Nitta:2012wi,Nitta:2012rq,Gudnason:2014nba}.
To conclude this result, the effective theory on a single soliton was used, 
which is nothing but a $d=2+1$ dimensional $U(1)$ gauged ${\mathbb C}P^1$ model without a potential in a constant magnetic field background.
In the previous studies \cite{Eto:2023lyo,Eto:2023wul,Eto:2023tuu}, 
we have used conventional BPS lumps
by neglecting the electromagnetic interaction 
as an approximation.
However, they should be replaced by gauged lumps constructed in this paper 
that may improve the phase boundary between the CSL and domain-wall Skyrmion phase.

This paper is organized as follows.
In Sec.~\ref{sec:nondynamical}, 
we construct gauged lumps with 
nondynamical gauge field in 
a constant magnetic field background.
In Sec.~\ref{sec:dynamical}, 
we consider dynamical gauge field 
and find that gauged lump solutions remain stable.
Sec.~\ref{sec:summary} is devoted to a summary and discussion.

\section{Gauged lumps with nondynamical gauge field in a constant magnetic field background}
\label{sec:nondynamical}

In this section, we consider nondynamical constant background gauge field. 
In Sec.~\ref{sec:model}, we briefly explain the gauged $\mathbb{C}P^1$ model. Then we explain the Derrick's theorem in Sec.~\ref{sec:derrick}, and show the numerical solutions for the single lump in Sec.~\ref{sec:numerical_sol} and for the higher charged lumps in Sec.~\ref{sec:numerical_sol_k}.

\subsection{The $U(1)$ gauged $\mathbb{C}P^1$ model}
\label{sec:model}

We consider a $U(1)$ gauged $O(3)$ model or equivalently $\mathbb{C}P^1$ model in $2+1$ dimensions. The Lagrangian is given by
\begin{eqnarray}
    {\cal L} = - \frac{1}{4}F_{\mu\nu}F^{\mu\nu} + D_\mu \vec n \cdot D^\mu \vec n\,,
    \label{eq:Lag}
\end{eqnarray}
where $\vec n$ is a three-component real column vector of scalar fields with the $S^2$ constraint
\begin{eqnarray}
    \vec n = \begin{pmatrix}
    n_1\\ n_2 \\ n_3
    \end{pmatrix}\,,\quad \vec n^2 = v^2\,,
\end{eqnarray}
where $v$ is a radius of $S^2$ whose mass dimension is $\dfrac{1}{2}$.
We gauge the $SO(2)$ symmetry about the $n_3$-axis as
\begin{eqnarray}
    n_1 + i n_2 \to e^{-i\theta(x)} \left(n_1 + i n_2\right)\,,\quad n_3 \to n_3\,,
\end{eqnarray}
with the covariant derivatives
\begin{eqnarray}
    D_\mu n_1 = \p_\mu n_1 + e A_\mu n_2\,,\quad
    D_\mu n_2 = \p_\mu n_2 - e A_\mu n_1\,,\quad
    D_\mu n_3 = \p_\mu n_3\,.    
\end{eqnarray}
Here, $A_\mu$ is the $U(1)$ gauge field and $e$ is the $U(1)$ gauge coupling constant. This can be also expressed as $D_\mu (n_1+in_2) = (\p_\mu - i e A_\mu)(n_1 + i n_2)$ with 
$A_\mu \to A_\mu - \p_\mu \theta / e$. The mass dimensions are summarised as 
$[\vec n] = [v] = \dfrac{1}{2}$,
$[A_\mu] = \dfrac{1}{2}$, and
$[e] = \dfrac{1}{2}$.
The symmetry of the model is $U(1) \times \mathbb{Z}_2$ where the $\mathbb{Z}_2$ is $n_3 \to -n_3$.


In the rest of this section we will consider a constant magnetic field background $F_{12} = B$, where the gauge field is not dynamical. In order to distinguish the nondynamical background gauge field from a dynamical one, we will use ${\cal A}_\mu$ for the former. Hereafter, we will take 
\begin{eqnarray}
    {\cal A}_0 = 0\,,\quad {\cal A}_1 = - \frac{By}{2}\,,\quad {\cal A}_2 = \frac{Bx}{2}\,.
    \label{eq:A_background}
\end{eqnarray}

\subsection{Derrick's theorem}
\label{sec:derrick}
We will study the topological lump solutions for $B\neq0$ below.
The static energy of the scalar field in the temporal gauge ${\cal A}_0=0$ is given by
\begin{align}
    E[\vec n] =\int d^2 x\,
     |D_i \Vec{n}|^2
    =\int d^2 x\left[
    |\partial_i \Vec{n}|^2+2e{\cal A}_i\varepsilon_{ab}\partial_in_an_b + e^2{\cal A}_i^2 (n_1^2+n_2^2)
    \right]\,.
\end{align}
Here we remove infinite energy of the constant magnetic field, so that $E[\vec n]$ can be 
finite.
We now want to apply the Derrick's scaling argument to $E[\vec n]$.  
For this purpose, we divide the energy into three pieces as
\begin{equation}
    E=E_2 + E_1 + E_0
\end{equation}
with 
\begin{eqnarray}
    E_2 &=& \int d^2 x\, |\partial_i \Vec{n}|^2 \, ,
    \\
    E_1 &=& \int d^2 x ~ 2e{\cal A}_i\varepsilon_{ab}(\partial_in_a)n_b
    \, , 
    \\
    E_0 &=& \int d^2x  ~ e^2{\cal A}_i^2 (n_1^2+n_2^2) \, .    
\end{eqnarray}
where the indices on $E$ represent the number of the partial derivative. The indices $a$ and $i$ runs from 1 to 2.
Note that $E_0$ and $E_2$ are positive semidefinite while $E_1$ can be either positive, negative or zero.

Let us define the scaling of $\vec n(x)$ as usual
\begin{eqnarray}
    \vec n^{(\lambda)}(x) = \vec n(\lambda x)\,,\quad
    \p_i \vec n^{(\lambda)}(x) = \lambda \p_i' \vec n(\lambda x)\,,
    \label{eq:scale_n}
\end{eqnarray}
where $\p_i'$ stands for $\p/\p (\lambda x^i)$.
On the other hand, the scaling of the gauge field differs from the dynamical  one because we deal with the constant magnetic field and the gauge field is not dynamical. We define the scaling of the background gauge field as follows
\begin{eqnarray}
    {\cal A}_i^{(\lambda)}(x) = \lambda^{-1} {\cal A}_i(\lambda x)\,.
    \label{eq:scale_A}
\end{eqnarray}
Plugging Eq.~(\ref{eq:A_background}) into this, we indeed see ${\cal A}_i^{(\lambda)}(x) = {\cal A}_i(x)$, and therefore
the constant background magnetic field is not affected by the rescaling.

The $\lambda$ dependence of the energy functional reads
\begin{eqnarray}
    e(\lambda) &=& E\left[\vec n^{(\lambda)}(x)\right] \nonumber\\
    &=& \int d^2 x\bigg[
    \left(\frac{\p n_a(\lambda x)}{\p x^i} \right)^2 +2e\left(\lambda^{-1}{\cal A}_i(\lambda x)\right)\varepsilon_{ab}\frac{\p n_a(\lambda x)}{\p x^i}n_b(\lambda x)\nonumber\\
    && + e^2\left(\lambda^{-1}{\cal A}_i(\lambda x)\right)^2 (n_1(\lambda x)^2+n_2(\lambda x)^2)
    \bigg]\nonumber\\
    &=& \int \frac{d^2x'}{\lambda^2} \bigg[
    \left(\lambda\frac{\p n_a(x')}{\p x'{}^i} \right)^2 +2e\left(\lambda^{-1}{\cal A}_i(x')\right)\varepsilon_{ab} \left(\lambda \frac{\p n_a(x')}{\p x'{}^i}\right) n_b(x')\nonumber\\
    && + e^2\left(\lambda^{-1}{\cal A}_i(x')\right)^2 (n_1(x')^2+n_2(x')^2)
    \bigg] \nonumber\\
    &=& E_2 + \lambda^{-2} E_1 + \lambda^{-4} E_0\,,
\end{eqnarray}
where we have used $x'{}^j = \lambda x^j$.
In order to see if this functional has a stationary point or not, we differentiate this with respect to $\lambda$ to obtain
\begin{eqnarray}
    \frac{d e(\lambda)}{d\lambda} = - 2\lambda^{-3}E_1 - 4 \lambda^{-5} E_0\,.
\end{eqnarray}
If $\vec{n}(x)$ is a static solution, this must be zero at $\lambda = 1$, provided
\begin{eqnarray}
    \delta \equiv E_1 + 2 E_0 = 0\,.
    \label{eq:Derrick}
\end{eqnarray}
As mentioned above $E_0 \ge 0$ whereas $E_1$ can be positive or negative. Hence, the system can successfully evade the Derrick's no-go theorem only for $E_1 < 0$.

\subsection{$k=\pm 1$ lump solution and the selecting rule}
\label{sec:numerical_sol}

Under the presence of nonzero background magnetic field $B\neq0$, the charged field cannot be condensed in the vacuum. Therefore, the vacuum configuration should be $n_3 = \pm v$ with $n_1+in_2=0$. Hence, the $\mathbb{Z}_2$ symmetry is spontaneously broken. We refer $n_3 = +v$ to the up vacuum while $n_3 = - v$ to the down vacuum.

Let us construct lump and anti-lump solutions.
To this end, we make an axially symmetric Ansatz, given by
\begin{eqnarray}
    n_1 =  v\frac{x}{r}\sin\Theta(r) \,,\quad
    n_2 = v\frac{y}{r}\sin\Theta(r) \,,\quad
    n_3 = v\cos\Theta(r)\,,
    \label{eq:ansatz_1lump}
\end{eqnarray}
with $r = \sqrt{x^2+y^2}$.
The lump with the winding number $k=+1$ satisfies the boundary condition $\Theta(0) = 0$ and $\Theta(\infty) = \pi$. Equivalently, this can be expressed as $n_3(0) = +v$ and $n_3(\infty) = -v$. Thus, this lump with $k=+1$ lives in the down vacuum $n_3 = - v$. Eq.~(\ref{eq:ansatz_1lump}) can also describe an anti-lump with $k=-1$ with the different boundary condition: $\Theta(0) = \pi$ and $\Theta(\infty) = 0$ or equivalently $n_3(0) = -v$ and $n_3(\infty) = +v$. Hence, the anti-lump lives in the up vacuum $n_3 = + v$. The lump in the down vacuum and the anti-lump in the up vacuum can be exchanged by $\Theta \to \pi - \Theta$ or $n_3 \to - n_3$.

Plugging this into Eq.~(\ref{eq:Lag}), we find the Lagrangian written in terms of $\Theta$ as
\begin{eqnarray}
    {\cal L} = - \frac{B^2}{2} - v^2\Theta'{}^2 - \frac{v^2(2 - eBr^2)^2}{4r^2}\sin^2\Theta\,,
    \label{eq:L_Theta}
\end{eqnarray}
and the corresponding EOM reads
\begin{eqnarray}
    \Theta'' + \frac{\Theta'}{r}  - \frac{(2 - eBr^2)^2}{4r^2}\sin\Theta\cos\Theta = 0\,,
\label{eq:eom_const_B}
\end{eqnarray}
where $v$ disappears from EOM.
Here, it is important to point out that EOM is not invariant under flipping the sign of $eB$. Note that the sign of $eB$ is also important in the Derrick's scaling condition.
Having the Ansatz (\ref{eq:ansatz_1lump}), we can express $E_0$ and $E_1$ with respect to $\Theta(r)$ as
\begin{eqnarray}
    E_0 &=& \frac{\pi e^2B^2v^2}{2}\int^\infty_0 dr\, r^3 \sin^2\Theta\,,\\
    E_1 &=& - 2\pi e B v^2 \int^\infty_0 dr\, r \sin^2\Theta\,.
    \label{eq:E1_1lump}
\end{eqnarray}
Both $E_0$ and $E_1$ can be nonzero for $B\neq0$, thus the above scaling argument is meaningful only for the lump under a nonzero magnetic field.
While $E_0 \ge 0$ is obvious from this expression, the sign of $E_1$ is the same as $-eB$. Since it has to be negative for Eq.~(\ref{eq:Derrick}) to be satisfied, we should choose $eB > 0$ for the Ansatz (\ref{eq:ansatz_1lump}).
More specifically, the lump in the down vacuum and the anti-lump in the up vacuum can only exist for $eB >0$.

We can take another Ansatz with $n_2$ being replaced by $n_2 = - v\dfrac{y}{r}\sin\Theta$ in Eq.~(\ref{eq:ansatz_1lump}) while $n_1$ and $n_3$ are unchanged. We call this transformation ($n_2 \to - n_2$ up to $U(1)$ gauge transformation) the topological charge (TC) conjugation. 
This should be distinguished from the previous transformation $n_3 \to - n_3$  that flips not only the winding number but also the vacuum. In contrast, the only lump charge is flipped by the TC conjugation. Indeed, the solution satisfying the boundary condition $\Theta(0) = 0$ and $\Theta(\infty) = \pi$ (equivalently $n_3(0) = +v$ and $n_3(\infty) = -v$) lives in the down vacuum and has the winding number $k=-1$, whereas that satisfying $\Theta(0) = \pi$ and $\Theta(\infty) = 0$ (equivalently $n_3(0) = -v$ and $n_3(\infty) = +v$) lives in the up vacuum and has the winding number $k=+1$.
Note that the TC conjugation is identical to the normal charge conjugation $n_1 + i n_2 \to n_1 - i n_2$. The charge conjugation is also same as $e \to -e$. Therefore, Eqs.~(\ref{eq:L_Theta})--(\ref{eq:E1_1lump}) remain correct with replacement $eB \to - eB$. Hence, we now need $eB < 0$ for realizing $E_1 < 0$. More specifically, the anti-lump in the down vacuum and the lump in the up vacuum can only exist for $eB < 0$.

\begin{figure}[t]
\begin{center}
\includegraphics[width=13cm]{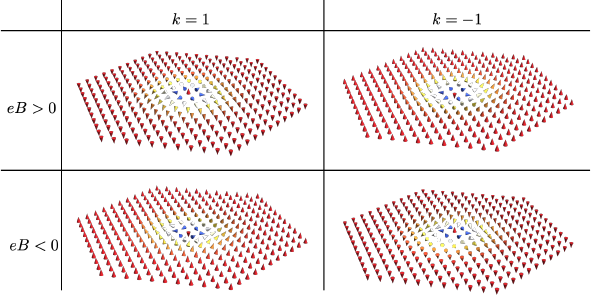}
\caption{The selecting rule. The cones stand for $\vec n$. The cone upward is $n_3 = +v$ and that downward is $n_3=-v$. The TC conjugation exchanges two diagonal entries, and two off-diagonal entries.}
\label{fig:lumps}
\end{center}
\end{figure}

\paragraph{The selecting rule and soliton-anti-soliton asymmetry}
This results in an interesting phenomenon. 
To clarify the lump solutions, we need to specify  signs for the following three quantities: the first is the sign for the vacuum ($n_3 = \pm v$), the second is that for the magnetic field ($eB >0$ or $eB <0$), and the third is that for the topological charge ($k>0$ or $k<0$). 
If the background magnetic field is absent, we can put a lump or an anti-lump either in the up or down vacuum. However, when $B \neq 0$,
there is a selecting rule that the lumps with positive topological charges ($k > 0$) can exist only when $\text{sign}(n_3)\cdot\text{sign}(eB) < 0$ is satisfied. Similarly, the anti-lumps with a negative topological charge ($k < 0$) can exist only when $\text{sign}(n_3)\cdot\text{sign}(eB) > 0$ holds. Hence, once the vacuum is chosen,  one can select either  lumps or anti-lumps by applying magnetic field. We call this phenomenon a violation of the TC conjugation, see Fig.~\ref{fig:lumps}. 

Solutions for $B\neq0$ can be only available by a numerical computation. However, asymptotic behavior can be analytically understood as follows. Since EOM includes $eB$ (its mass dimension is $2$), a solution will not have a size modulus. Indeed, the  asymptotic behavior should obey an exponential law 
rather than a power law. This can be easily verified by analyzing the linearized EOM for the fluctuation
$\vartheta(r) = \pi - \Theta(r)$ at $r \gg 1/\sqrt{eB}$ as
\begin{eqnarray}
    - \vartheta'' + \frac{e^2B^2r^2}{4}\vartheta = 0\,,\quad
    \vartheta \propto \exp\left(-\frac{eB}{4}r^2\right)\,,
\end{eqnarray}
with the expected decay length $1/\sqrt{eB}$.

\begin{figure}[t]
\centering
\begin{minipage}[t]{0.49\linewidth}
\centering
\includegraphics[keepaspectratio,scale=0.4]{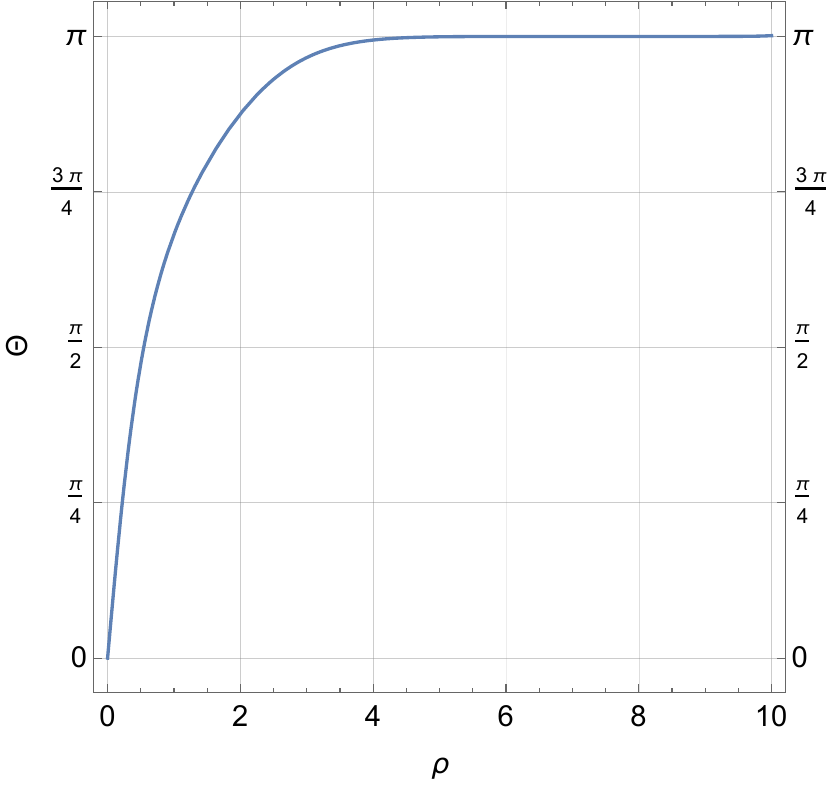}
\end{minipage}
\begin{minipage}[t]{0.49\linewidth}
\centering
\includegraphics[keepaspectratio,scale=0.38]{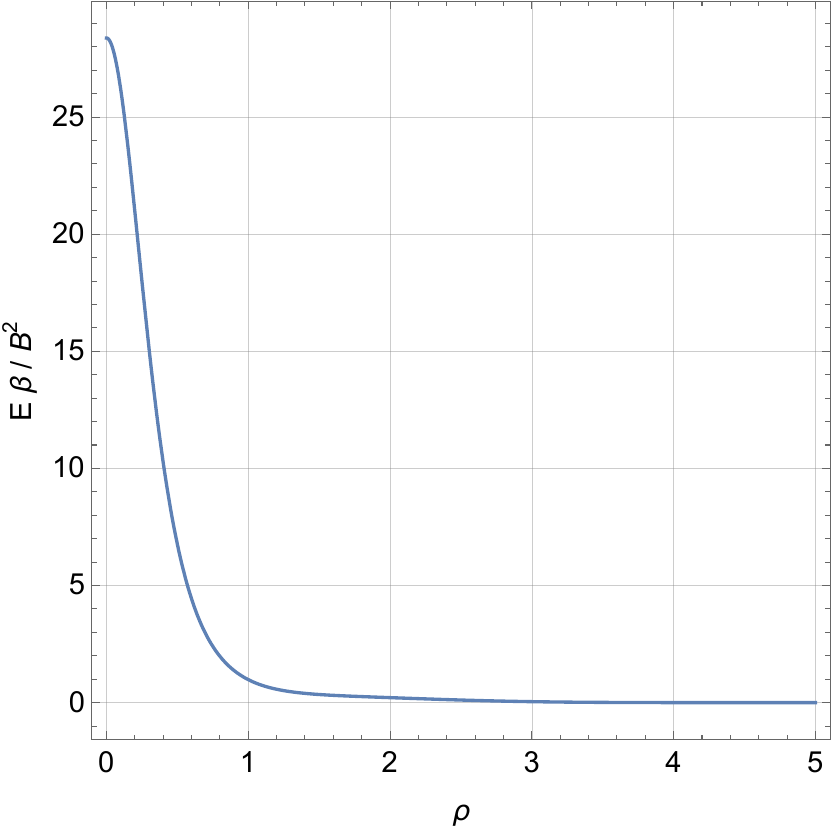}
\end{minipage}
\caption{The numerical solution $\Theta$ and energy density of the single lump for $eB>0$.}
\label{fig:solutions_B_const}
\end{figure}

Let us show the numerical solution. To this end, we first change the variable as $\rho = \sqrt{eB}\,r$.
Then, EOM to be solved uniquely becomes  
\begin{eqnarray}
    \Theta'' + \frac{\Theta'}{\rho}  - \frac{(2 - \rho^2)^2}{4\rho^2}\sin\Theta\cos\Theta = 0\,,
\label{eq:eom_const_B_rescale}
\end{eqnarray}
where $\Theta' = \dfrac{d\Theta}{d\rho}$.
The energy density and the mass are expressed as
\begin{eqnarray}
    {\cal E} =  
    \frac{B^2}{\beta} \left[ \Theta'{}^2 +  \frac{\left(2- \rho^2\right)^2}{4\rho^2} \sin^2\Theta\right]\,,\quad
    M = 2\pi v^2 \int \rho d\rho\, {\cal E}\,,
\end{eqnarray}
respectively.
Here, we have introduced the dimensionless parameter for later convenience
\begin{eqnarray}
    \beta \equiv \frac{B}{ev^2}\,.
    \label{eq:beta}
\end{eqnarray}
A numerical solution is shown in Fig.~\ref{fig:solutions_B_const}.
Let us discuss features of this solution. 
First, we verify the Derrick's constraint in Eq.~(\ref{eq:Derrick}). With respect to the dimensionless coordinate, we have
\begin{eqnarray}
    E_0 &=& 2\pi v^2 \int^\infty_0 d\rho\, \frac{\rho^3}{4} \sin^2\Theta\,,\\
    E_1 &=& - 2\pi v^2 \int^\infty_0 d\rho\, \rho \sin^2\Theta\,,
\end{eqnarray}
and we numerically obtain $\delta/(2\pi v^2) = 3.32494 \times 10^{-7}$, thus Eq.~(\ref{eq:Derrick}) is satisfactory met within numerical accuracy. 
The mass of the lump is numerically evaluated as $M/(2\pi v^2) = 3.62188$. Note that this does not depend on $eB$.
We also numerically determine the size $r_0 = \dfrac{0.544784}{\sqrt{eB}}$ of the lump from the condition $n_3(r_0)=0$ $\left(\Theta(r_0)=\dfrac{\pi}{2}\right)$. Hence, the magnetic flux inside the lump can be evaluated as $\Phi = \pi r_0^2 B \simeq 0.15 \times \dfrac{2\pi}{e}$.

We can compare our numerical solution with the BPS lump at $B=0$, see Appendix \ref{apndx:BPS_lump} for some details. The mass of BPS lump is $M_\text{BPS}/(2\pi v^2) = 4$,
therefore our numerical solution for $B\neq 0$ is slightly lighter than the BPS lump at $B=0$.
Furthermore, rough estimation of the fixed size of a BPS lump under the constant magnetic field $B$ was given as $\tilde r_0 = \sqrt{\dfrac{2}{eB}}$ due to the magnetic flux quantization condition that the magnetic flux inside $\tilde r_0$ is $\tilde\Phi = \dfrac{2\pi}{e}$ \cite{Eto:2023lyo}.
Our numerical solution improves this, and the actual size is about 40\% of $\tilde r_0$, and the magnetic flux inside the lump is about 15\% of $\tilde \Phi$.
 
\subsection{Coaxial multiple lumps}
\label{sec:numerical_sol_k}

\begin{figure}[t]
\centering
\begin{minipage}[t]{0.49\linewidth}
\centering
\includegraphics[keepaspectratio,scale=0.4]{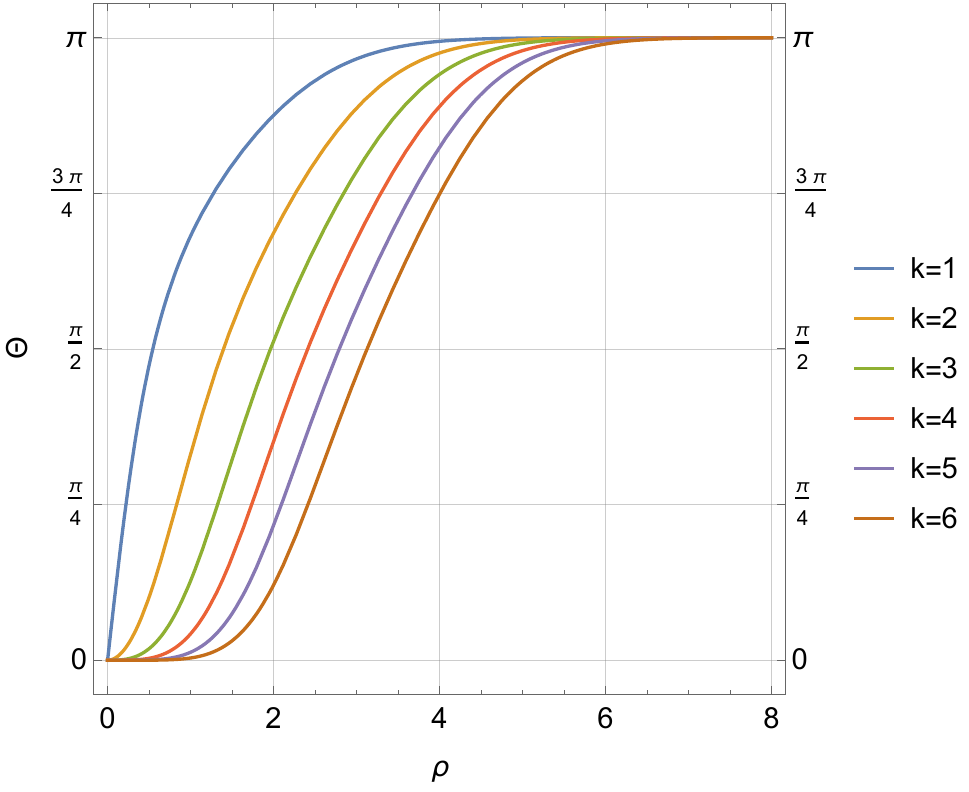}
\end{minipage}
\begin{minipage}[t]{0.49\linewidth}
\centering
\includegraphics[keepaspectratio,scale=0.38]{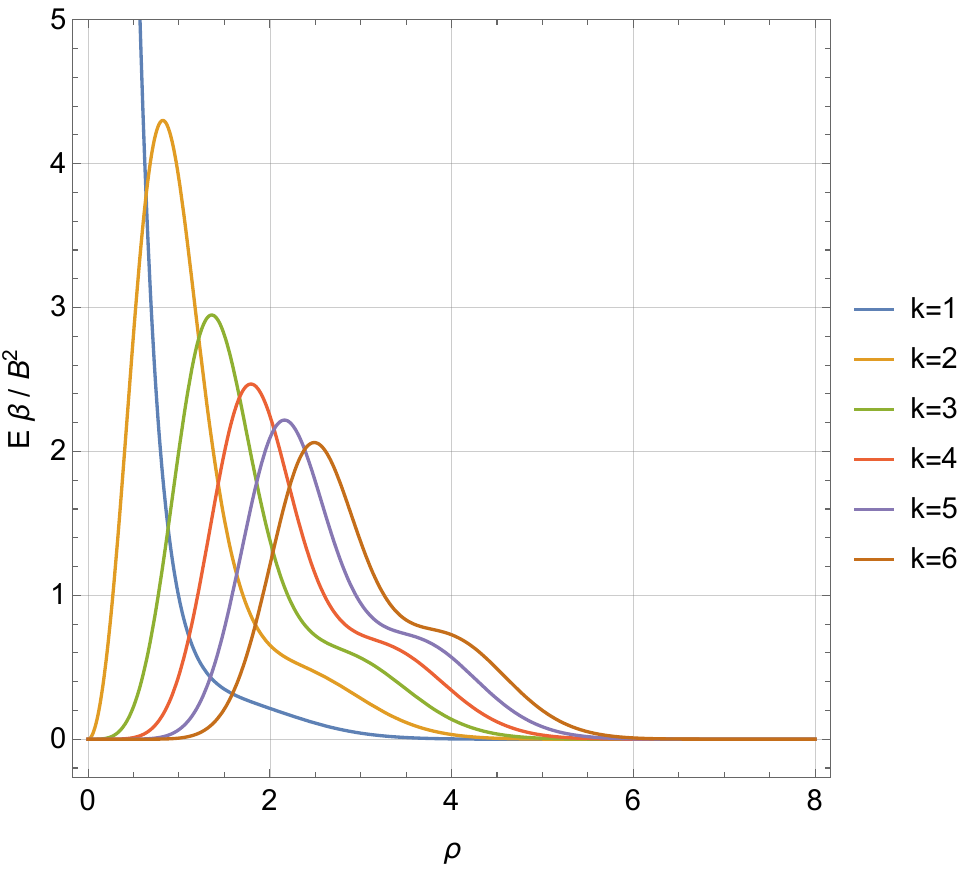}
\end{minipage}
\caption{The numerical solutions $\Theta$ and energy densities for $k=1,2,3,4,5$ with $eB > 0$.}
\label{fig:solutions_B_const_k}
\end{figure}

We here quickly show the axially symmetric lumps with the winding number $k>1$. To this end, we slightly change the Ansatz
\begin{eqnarray}
    n_1 =  v\frac{{\rm Re}\,z^k}{r^k}\sin\Theta(r) \,,\quad
    n_2 = v\frac{{\rm Im}\,z^k}{r^k}\sin\Theta(r) \,,\quad
    n_3 = v\cos\Theta(r)\,,
    \label{eq:ansatz_k_lump}
\end{eqnarray}
with $z \equiv x+ iy$.
The EOM reads
\begin{eqnarray}
    \Theta'' + \frac{\Theta'}{r}  - \frac{(2k - eBr^2)^2}{4r^2}\sin\Theta\cos\Theta = 0\,,
\label{eq:eom_const_B_k}
\end{eqnarray}
and $E_0$ and $E_1$ are expressed as
\begin{eqnarray}
    E_0 &=& 2\pi v^2 \int^\infty_0 d\rho\, \frac{\rho^3}{4} \sin^2\Theta\,,\\
    E_1 &=& - 2k \pi v^2 \int^\infty_0 d\rho\, \rho \sin^2\Theta\,.
    \label{eq:E1_1lump_k}
\end{eqnarray}
Here $E_0$ is independent of $k$ while $E_1$ is proportional to $k$.
The numerical solutions for $k=1,2,3,4,5,6$ are shown in Fig.~\ref{fig:solutions_B_const_k}.
We numerically verify if the Derrick's condition is satisfied and find $\delta/(2\pi v^2) = -6.83875 \times 10^{-7}$, $-1.17859\times 10^{-6}$, $-1.58112\times 10^{-6}$, $-1.63619 \times 10^{-6}$, $-2.50128 \times 10^{-6}$ for $k=2,3,4,5,6$, respectively. 

\begin{figure}[h]
\centering
\begin{minipage}[t]{0.49\linewidth}
\centering
\includegraphics[keepaspectratio,scale=0.45]{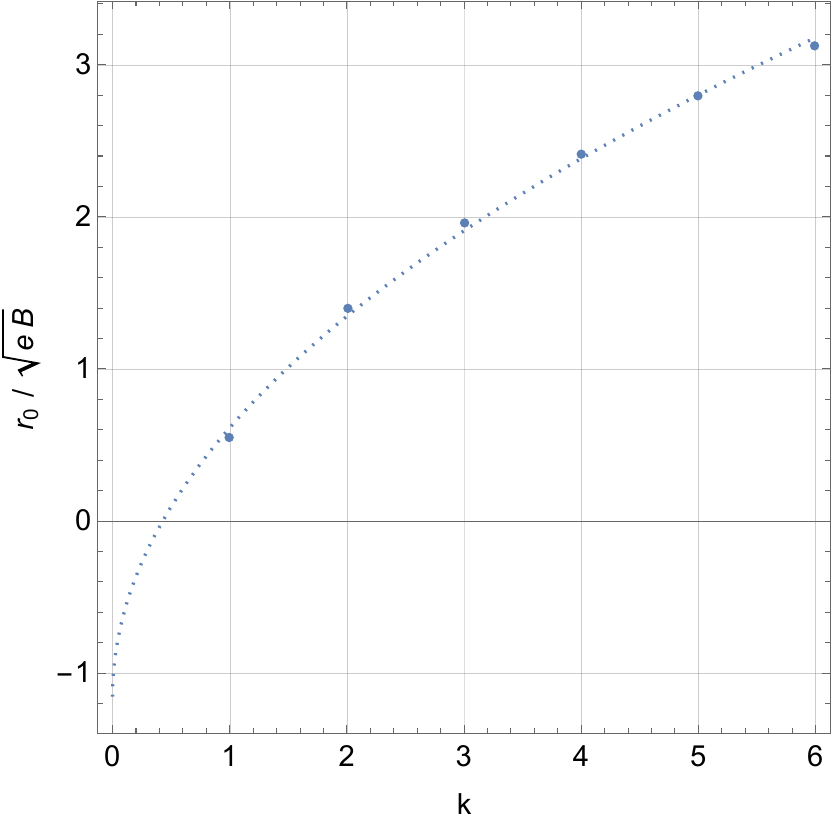}
\end{minipage}
\begin{minipage}[t]{0.49\linewidth}
\centering
\includegraphics[keepaspectratio,scale=0.45]{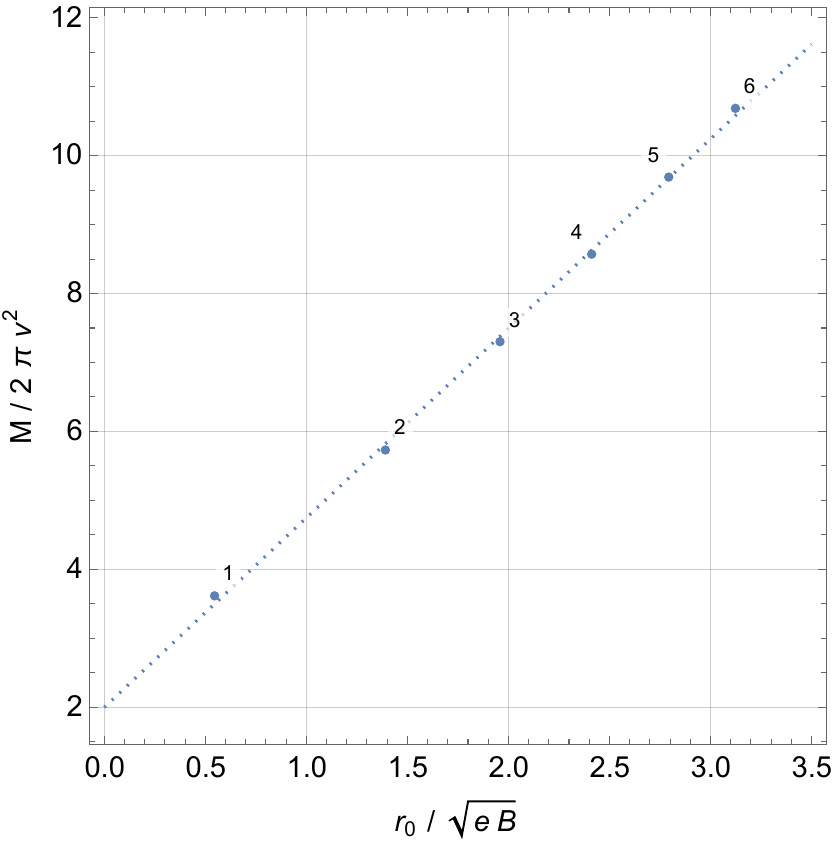}
\end{minipage}
\caption{The numerical solutions $\Theta$ and energy densities for $k=1,2,3,4,5,6$ with $eB > 0$.}
\label{fig:high_k}
\end{figure}

The lumps get fat as $k$ increased.
We numerically measure the radius $r_0$ for $k$ as shown in Fig.~\ref{fig:high_k} and find that it can be well approximated by $\dfrac{r_0}{\sqrt{eB}} = a \sqrt{k} + b$ with $a=1.77$ and $b=-1.16$. Similarly, the mass is found as $\dfrac{M}{2\pi v^2} = c \sqrt{k} + d$ with $c=4.86$ and $d=-1.18$. This suggests that the lumps are well described by a droplet model.
Note that the lumps with $k > 1$ are empty in the sense that the energy densities at the cores are negligibly small.  

\section{Gauged  lumps with dynamical gauge field 
in a magnetic field background}
\label{sec:dynamical}

In this section, we consider a dynamical gauge field with a uniform background magnetic field. 
In Sec.~\ref{sec:model_dyn}, we introduce the dynamical gauge field. Then we explain the Derrick's theorem in Sec.~\ref{sec:derrick_dyn}, and show the numerical solutions for the gauged lumps in Sec.~\ref{sec:numerical_sol_dyn}.
The anti-lumps and the soliton-anti-soliton asymmetry are explained in Sec.~\ref{sec:dyn_gauge_antilump}.

\subsection{The model with dynamical gauge field}
\label{sec:model_dyn}

Next we deal with $A_\mu$ as a dynamical field. Since we are interested in the lumps under the constant magnetic field, we first need to decompose the gauge field into two parts as
\be
A_\mu = {\cal A}_\mu + a_\mu\,,
\label{eq:A_separate}
\ee
where ${\cal A}_\mu$ is the nondynamical background gauge field given in Eq.~(\ref{eq:A_background}) and $a_\mu$ is the dynamical gauge field.

\subsection{Derrick's theorem}
\label{sec:derrick_dyn}

We modify the Derrick's scaling argument in the previous section to the case with a dynamical gauge field.
The static and finite energies of the scalar field in the temporal gauge $A_0=0$ are given by
\begin{eqnarray}
    E[\vec n,a_\mu] &=& \int d^2 x\,
    \left(
     \frac{F_{12}^2}{2} - \frac{B^2}{2} + |D_i \Vec{n}|^2
     \right) \nonumber\\
    &=& \int d^2 x\bigg[
    \frac{(B + f_{12})^2}{2}-\frac{B^2}{2}\nonumber\\
    &&+ |\partial_i \Vec{n}|^2+2e({\cal A}_i+a_i)\varepsilon_{ab}\partial_in_an_b + e^2({\cal A}_i+a_i)^2 (n_1^2+n_2^2)
    \bigg]\,,
\end{eqnarray}
respectively. 
We have subtracted $B^2/2$, and so that the energy remains finite. Let us decompose this into four pats as follows:
\begin{eqnarray}
    E[\vec n,a_\mu] = E_4 + E_2 + E_1 + E_0\,,
\end{eqnarray}
with
\begin{eqnarray}
E_4 &=& \int d^2x\, \frac{f_{12}^2}{2}\,,\\
E_2 &=& \int d^2x\, \left[B f_{12} + |\partial_i \Vec{n}|^2 +2e a_i\varepsilon_{ab}\partial_in_an_b + e^2 a_i^2 (n_1^2+n_2^2)\right]\,,\\
E_1 &=& \int d^2x\, \left[ 2e{\cal A}_i\varepsilon_{ab}\partial_in_an_b + 2 e^2 {\cal A}_i a_i (n_1^2 + n_2^2)
\right]\,,\\
E_0 &=& \int d^2x\, e^2{\cal A}_i^2 (n_1^2+n_2^2)\,,
\end{eqnarray}
and $f_{12} = \p_1 a_2 - \p_2 a_1$.

Now we apply the Derrick's scaling argument to these as before. To this end, we use the same scaling laws Eqs.~(\ref{eq:scale_n}) and (\ref{eq:scale_A}) for $\vec n$ and ${\cal A}_i$, respectively. On the other hand, we adopt the standard scaling law for the dynamical gauge field $a_\mu$ as
\begin{eqnarray}
    a_i^{(\lambda)}(x) = \lambda\, a_i(\lambda x)\,.
\end{eqnarray}
The $\lambda$ dependence of the energy functional reads
\begin{eqnarray}
    e(\lambda) = E[\vec n^{(\lambda)},a_\mu^{(\lambda)}]
    = \lambda^2 E_4 + \lambda^0 E_2 + \lambda^{-2}E_1 + \lambda^{-4}E_0\,.    
\end{eqnarray}
In order to see if this functional has a stationary point or not, we differentiate this with respect to $\lambda$ as
\begin{eqnarray}
    \frac{d e(\lambda)}{d\lambda} = 2\lambda E_4 - 2\lambda^{-3}E_1 - 4 \lambda^{-5} E_0\,.
\end{eqnarray}
If $\{\vec{n}(x),a_\mu\}$ is a static solution, this must be zero at $\lambda = 1$, provided
\begin{eqnarray}
    \delta' \equiv - E_4 + E_1 + 2 E_0 = 0\,.
    \label{eq:Derrick2}
\end{eqnarray}
As before, $E_1$ can be either positive or negative whereas $E_0$ and $E_4$ are positive.

\subsection{Multiple gauged lumps in the down vacuum with $eB > 0$}
\label{sec:numerical_sol_dyn}

For a single lump, 
we make the following Ansatz
\begin{eqnarray}
    a_0 = 0\,,\quad
    a_1 = - \frac{B a(r)}{2} y\,,\quad
    a_2 = \frac{B a(r)}{2} x\,,
\end{eqnarray}
where $a_\mu$ is the dynamical gauge field as  given in Eq.~(\ref{eq:A_separate}).
Together with the background gauge field in Eq.~(\ref{eq:A_background}), the full gauge field is given by
\begin{eqnarray}
    A_0 = 0\,,\quad
    A_1 = - \frac{B (1+a(r))}{2} y\,,\quad
    A_2 = \frac{B (1+a(r))}{2} x\,.
    \label{eq:ansatz_gauge}
\end{eqnarray}
The mass dimension of $a(r)$ is zero.
The magnetic field is given by
\begin{eqnarray}
    F_{12} = B (1+a) + \frac{B r a'}{2}\,.
\end{eqnarray}
We impose the profile function $a(r)$ to approach $0$ as $r \to \infty$, so that the magnetic field asymptotically behaves as $F_{12} \to B$.

Plugging Eqs.~(\ref{eq:ansatz_k_lump}) and (\ref{eq:ansatz_gauge}) into Eq.~(\ref{eq:Lag}), we find the reduced Lagrangian for $\Theta$ and $a$ 
\begin{eqnarray}
    {\cal L} = - \frac{B^2}{2}\left(1+a + \frac{r a'}{2}\right)^2 - v^2 \Theta'{}^2 - v^2 \frac{\left(2k- e B r^2 (1+a)\right)^2}{4r^2} \sin^2\Theta\,.
\end{eqnarray}
The corresponding EOMs are given by
\begin{eqnarray}
    &&\Theta'' + \frac{\Theta'}{r} - \frac{\left(2k-eBr^2(1+a)\right)^2}{4r^2}\sin\Theta\cos\Theta = 0\,,\\
&&a'' + \frac{3a'}{r}  + \frac{2 ev^2\left(2k - eBr^2 (1+a)\right)}{Br^2} \sin^2\Theta = 0\,.
\end{eqnarray}
Up to here in this subsection, the prime stands for a derivative in terms of the physical coordinate $r$.

For numerical analysis, let us rewrite these with respect to the dimensionless coordinate $\rho = \sqrt{eB}\,r$. Then we have
\begin{eqnarray}
    &&\Theta'' + \frac{\Theta'}{\rho} - \frac{\left(2k-\rho^2(1+a)\right)^2}{4\rho^2}\sin\Theta\cos\Theta = 0\,,\label{eq:eom_full_1}\\
    && a'' + \frac{3a'}{\rho}  +  \frac{2\left(2k - \rho^2 (1+a)\right)}{\beta\rho^2} \sin^2\Theta = 0\,.\label{eq:eom_full_2}
\end{eqnarray}
Now the prime indicates a derivative in terms of $\rho$.
The EOMs include the dimensionless parameter $\beta$ defined in Eq.~(\ref{eq:beta}).
If we assume $a=0$, Eq.~(\ref{eq:eom_full_1}) is identical to Eq.~(\ref{eq:eom_const_B_rescale}). On the other hand, Eq.~(\ref{eq:eom_full_2}) with $a=0$ cannot be satisfied except for $\Theta = 0$ or $\pi$. Eq.~(\ref{eq:eom_full_2}) with $a=0$ is approximately satisfied for a large $B$ (or small $e$) such as $B \gg ev^2$. This is a reasonable condition for us to ignore the back reaction to the background gauge field.

\begin{figure}[h]
\centering
\begin{minipage}[t]{0.49\linewidth}
\centering
\includegraphics[keepaspectratio,scale=0.4]{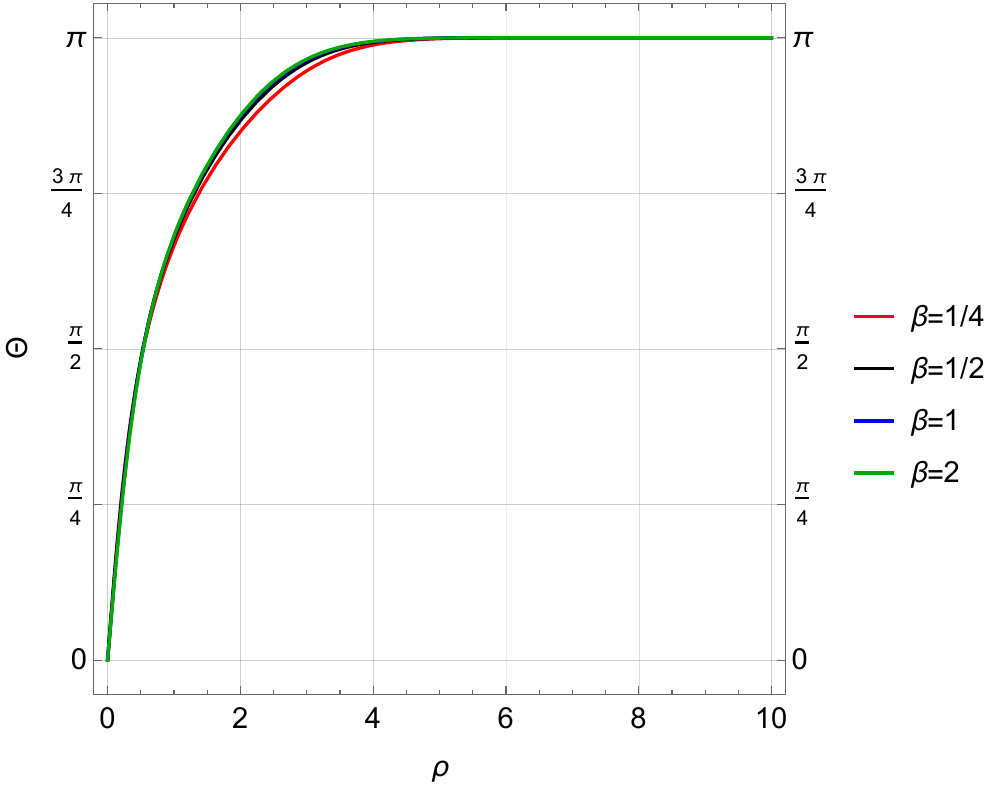}
\end{minipage}
\begin{minipage}[t]{0.49\linewidth}
\centering
\includegraphics[keepaspectratio,scale=0.37]{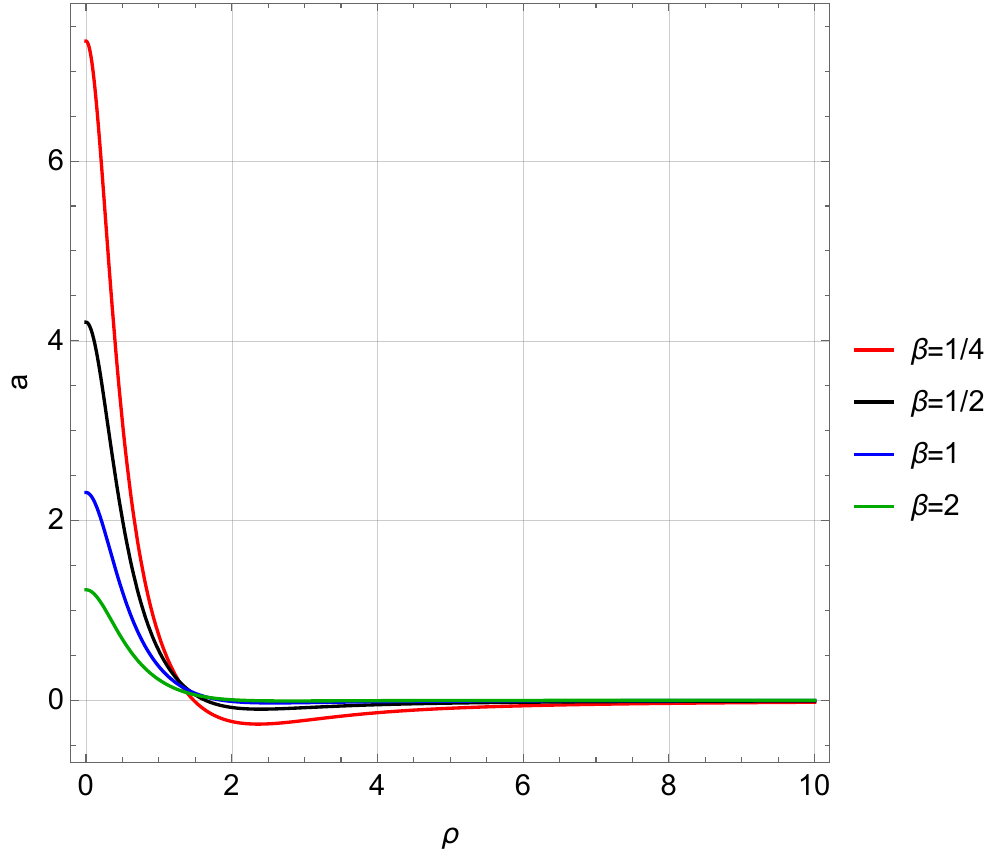}
\end{minipage}
\begin{minipage}[t]{0.49\linewidth}
\centering
\includegraphics[keepaspectratio,scale=0.37]{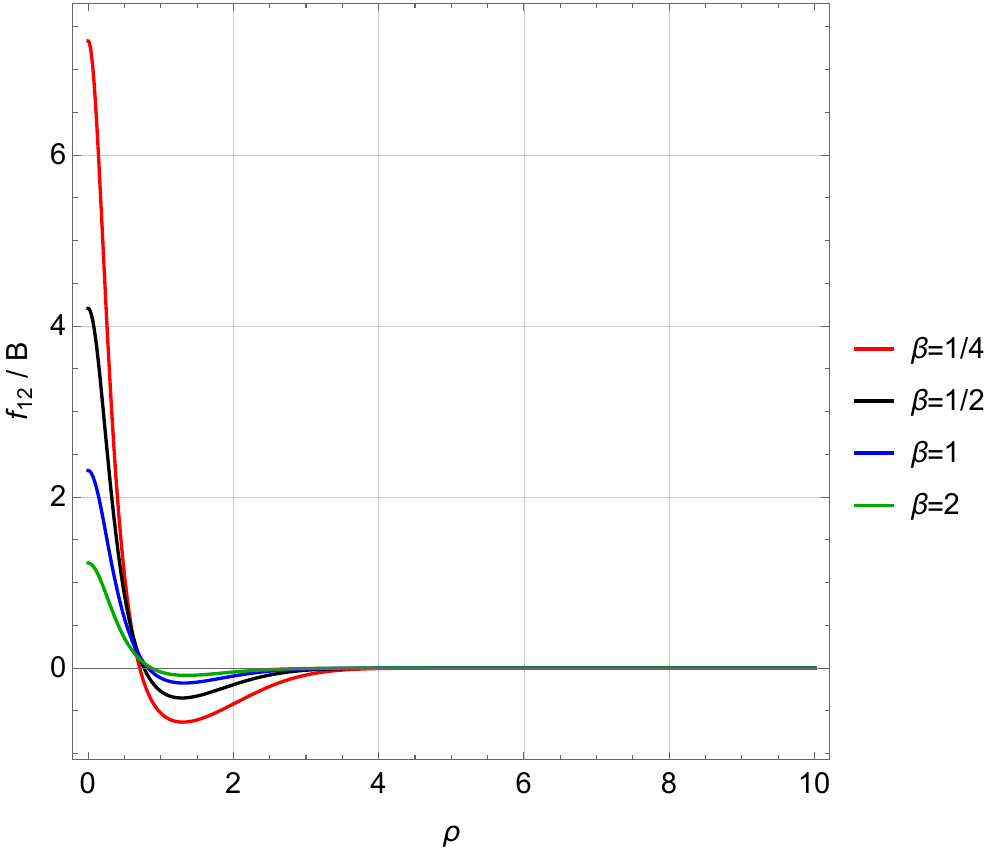}
\end{minipage}
\begin{minipage}[t]{0.49\linewidth}
\centering
\includegraphics[keepaspectratio,scale=0.38]{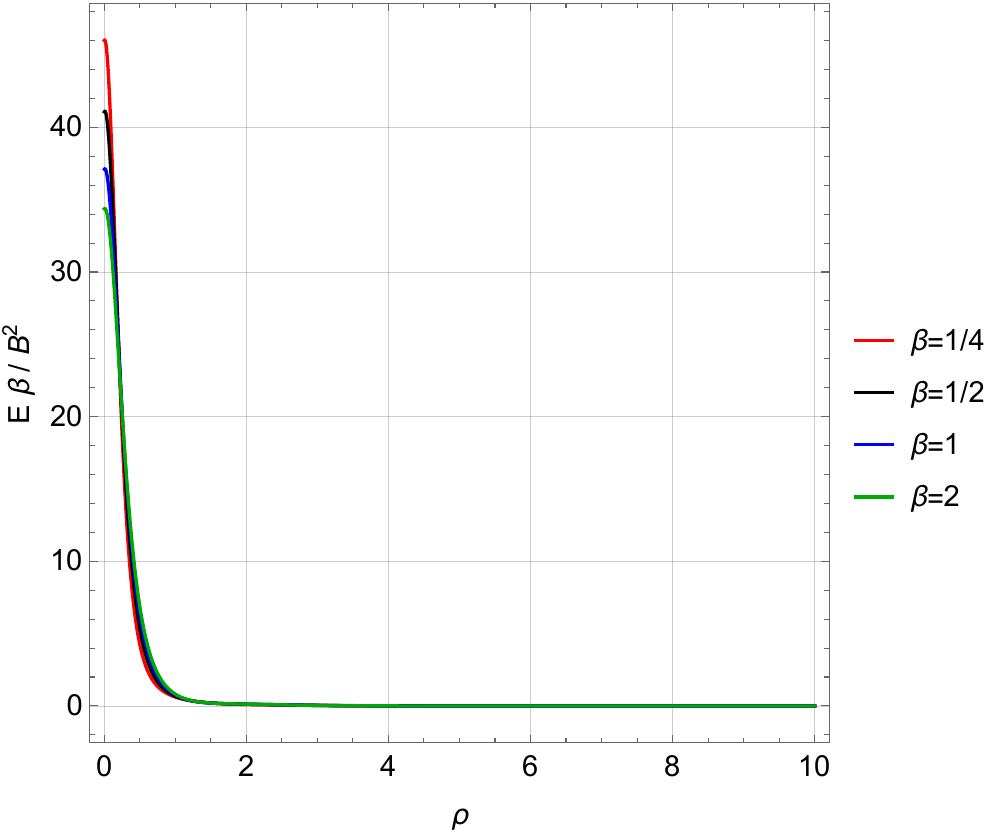}
\end{minipage}
\caption{The numerical solution of $k=1$ lump. $\Theta$ (top-left) and $a$ (top-right). The back reaction $f_{12}$ to the background magnetic field $B$ (bottom-left) and energy density excited by the single lump. We take $\beta = \{1/4,\,1/2,\,1,\,2\}$.}
\label{fig:solutions_B_dynamical}
\end{figure}

In what follows, we will concentrate on the lumps with $k>0$ in the down vacuum with $eB > 0$. Namely, the boundary condition for $\Theta$ is $\Theta(0)=0$ and $\Theta(\infty)=\pi$.
The numerical solutions of $k=1$ with different $\beta = \frac{1}{4}$, $\frac{1}{2}$, $1$, and $2$ are shown in Fig.~\ref{fig:solutions_B_dynamical}.
The anti-lumps with $k<0$ in the negative vacuum with $eB > 0$ will separately be explained in Sec.~\ref{sec:dyn_gauge_antilump}.
We first realize that back reaction to $\Theta$ is negligibly small irrespective of $\beta$, see the top-left panel. On the other hand, as can be seen in the top-right panel, the dynamical gauge field $a_\mu$ sensitively reacts to $\beta$.
We also show the back reactions of the magnetic flux density and energy density 
\begin{eqnarray}
    f_{12} &=& B \left(a + \frac{\rho}{2}a'\right)\,,\\
    {\cal E} &=& \frac{B^2}{\beta} \left[\frac{\beta}{2}\left\{\left(1 + a + \frac{\rho}{2}a'\right)^2 - \frac{1}{2}\right\} + \Theta'{}^2 + \frac{(2k-\rho^2(1+a))^2}{4\rho^2} \sin^2\Theta\right] \,,  
\end{eqnarray}
respectively, in the second row of Fig.~\ref{fig:solutions_B_dynamical}.
Note that the net magnetic flux is $F_{12} = B + f_{12}$ and we show $f_{12}$ in Fig.~\ref{fig:solutions_B_dynamical}. The magnetic flux is condensed at the center of the lump. It gets stronger than the background value $B$ inside the lump core, and it reduces and changes its sign  as going away from the lump, and asymptotically decays at spatial infinity.
As expected, the greater $\beta$ is, the smaller deformation of the profile $a$ is. Namely,
the back reaction tends to be small for the strong background magnetic field $B \gg ev^2$.
This behavior can be explained as follows. Since $|n_1+in_2| \neq 0$ at the edge of lump, the lump behaves as a superconducting ring 
and traps magnetic field in its inside, 
with the total magnetic field reduced from 
the background magnetic field because of
superconductiviy expelling it.

Let us quantitatively evaluate the numerical solutions. First, we measure a net change of the magnetic flux due to the single lump,
\begin{eqnarray}
    \delta \Phi 
    = \int d^2x\, f_{12}
    = \frac{2\pi}{e} \int \rho d\rho\,\left(a+\frac{\rho}{2}a'\right)\,.
\end{eqnarray}
This is shown 
for $\beta = \{\frac{1}{4},\,\frac{1}{2},\,1,\,2\}$ 
in Table \ref{table}.
We find that $\delta \Phi$ is always negative, and thus the net background magnetic field is weakened under the presence of lump regardless of $\beta$.

Second, we measure the mass of the lump by
\begin{eqnarray}
    M 
    = 2\pi v^2 \int \rho d\rho\, \left[\frac{\beta}{2}\left\{\left(1 + a + \frac{\rho}{2}a'\right)^2 - 1\right\} + \left\{\Theta'{}^2 + \frac{(2k-\rho^2(1+a))^2}{4\rho^2} \sin^2\Theta\right\}\right]\,.
\end{eqnarray}
This is also summarised in Table \ref{table}.

Third,
we calculate $E_4$, $E_1$, and $E_0$ by
\begin{eqnarray}
    E_4 &=& \frac{2\pi B}{e} \int \rho d\rho\,  \frac{1}{2}\left(a + \frac{\rho}{2}a'\right)^2\,,\\
    E_1 &=& \frac{2\pi B}{e\beta} \int \rho d\rho\,  \left(\frac{\rho^2 a}{2}-k\right)\sin^2\Theta\,,\\
    E_0 &=& \frac{2\pi B}{e\beta} \int \rho d\rho\, \frac{\rho^2}{4} \sin^2\Theta\,,
\end{eqnarray}
and verify the scaling-stability condition $\delta' =0$.
The results are summarize in Table.~\ref{table} and they are satisfactory small.

\begin{table}[t]
\begin{center}
\begin{tabular}{|c|cccc|c|}
\hline
 $k=1$ & $\beta=1/4$ & $\beta = 1/2$ & $\beta = 1$ & $\beta = 2$ & $\beta=\infty$\\
\hline
$\delta \Phi/(2\pi/e)$ & $-1.11$ & $-0.400$ & $-0.132$ & $-0.0402$ & $0$\\
$M/(2\pi v^2)$ & $2.90$ & $3.14$ & $3.33$ & $3.45$ & $3.62$\\
$\delta'/(2\pi B/e)$ & $-5.61 \times 10^{-4}$
& $-7.18\times 10^{-4}$  
& $-2.43\times10^{-4}$  
& $-8.84\times10^{-5}$  
& ---\\
\hline
\end{tabular}
\caption{The quantitative properties of the $k=1$ lump for $\beta = 1/2$, $1$, and $2$. We numerically evaluate changes of the net magnetic flux, the mass, and the Derrick's scaling condition $\delta' = 0$.}
\label{table}
\end{center}
\end{table}

\begin{figure}[t]
\centering
\begin{minipage}[t]{0.49\linewidth}
\centering
\includegraphics[keepaspectratio,scale=0.4]{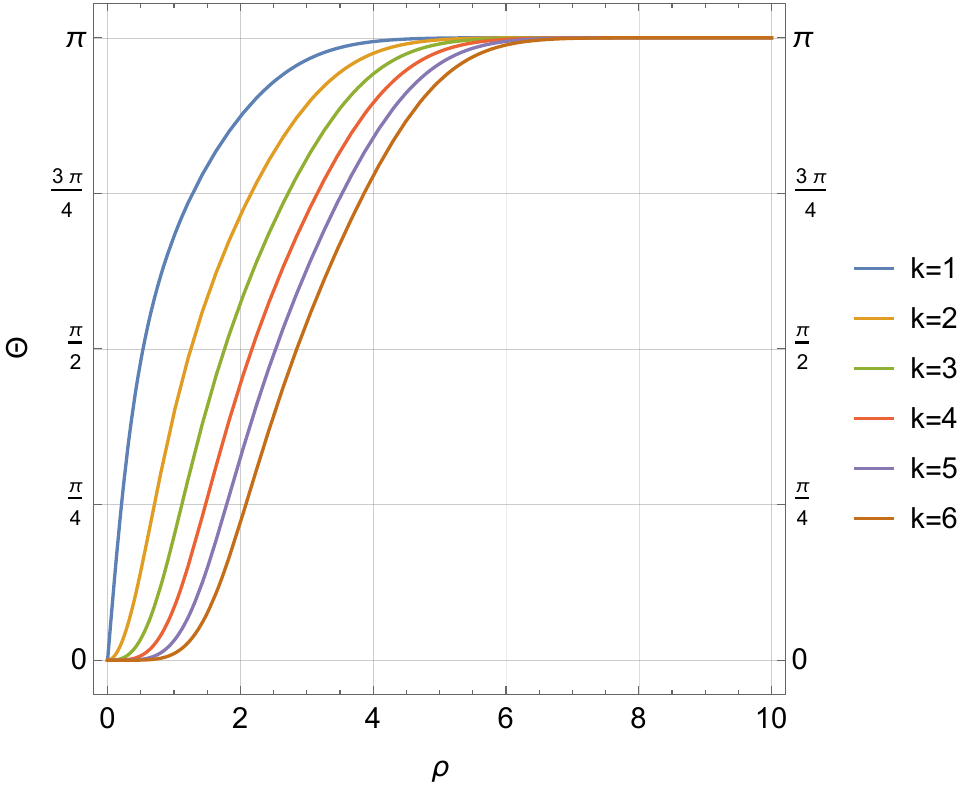}
\end{minipage}
\begin{minipage}[t]{0.49\linewidth}
\centering
\includegraphics[keepaspectratio,scale=0.4]{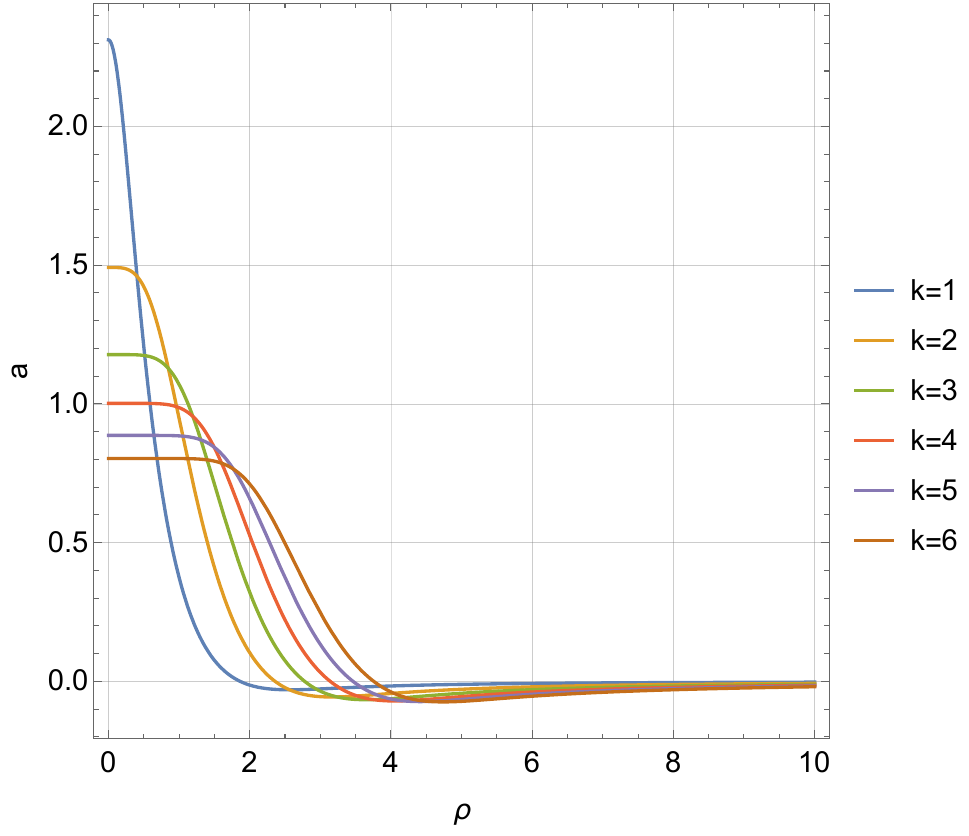}
\end{minipage}
\begin{minipage}[t]{0.49\linewidth}
\centering
\includegraphics[keepaspectratio,scale=0.4]{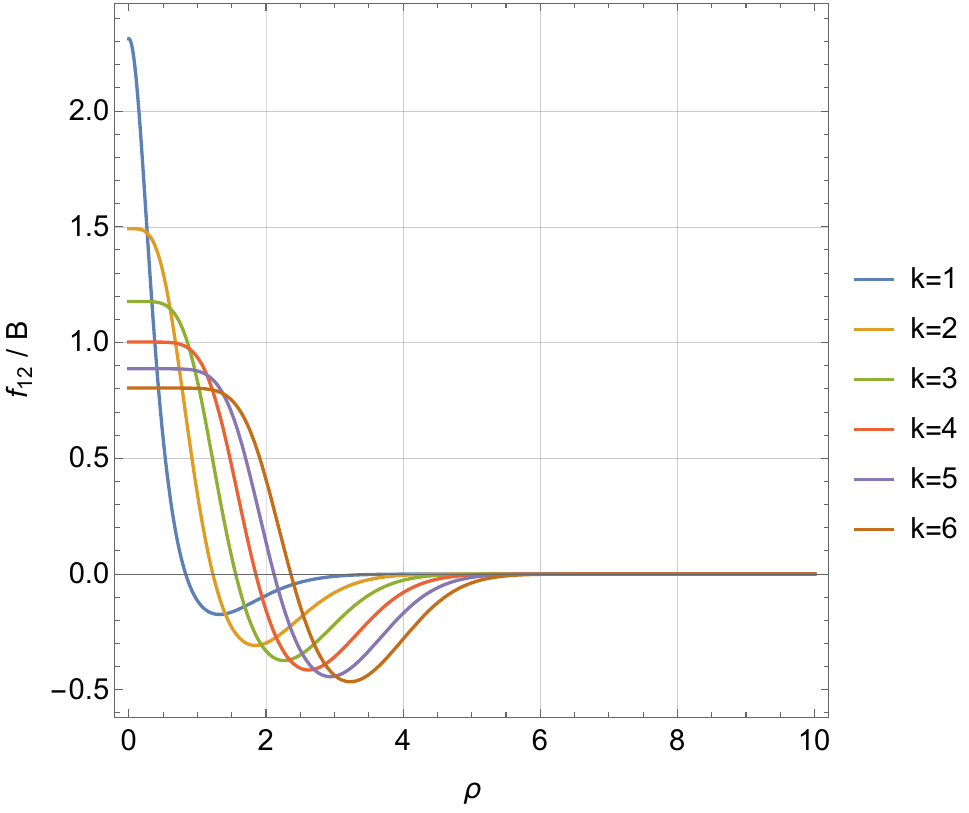}
\end{minipage}
\begin{minipage}[t]{0.49\linewidth}
\centering
\includegraphics[keepaspectratio,scale=0.4]{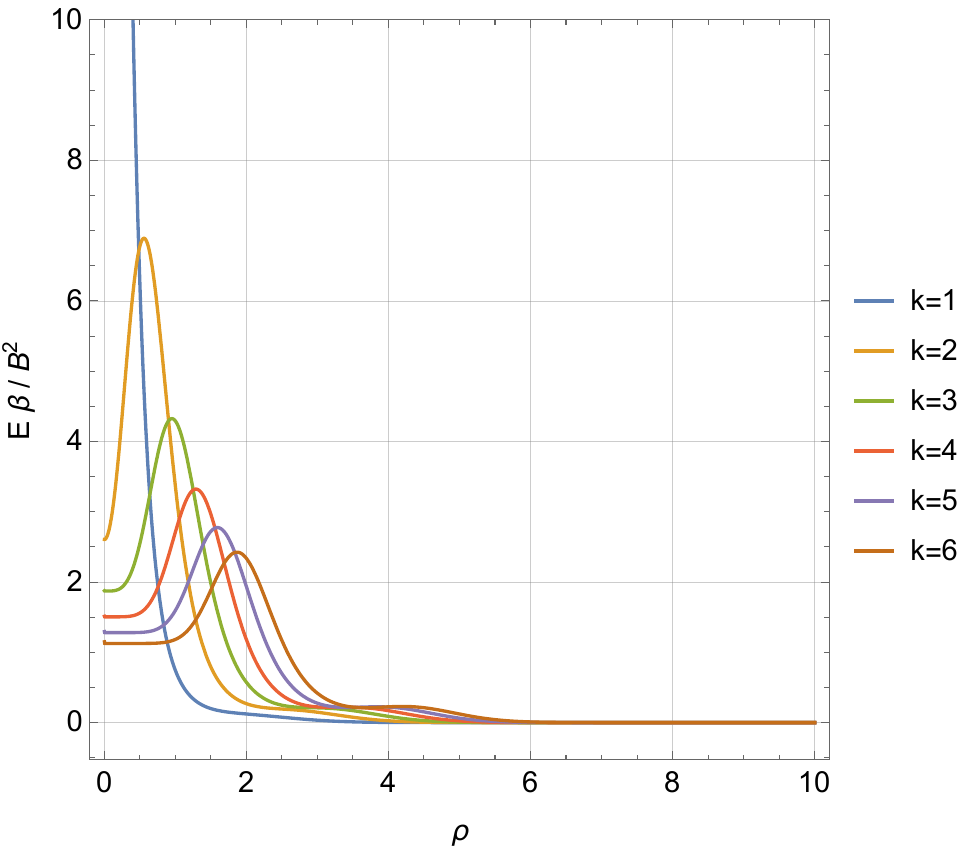}
\end{minipage}
\caption{The axially symmetric lumps for $k=1,2,3,4,5,6$ for $\beta=1$. $\Theta$ and $a$ are shown in the top row, and $f_{12}$ and ${\cal E}$ are shown in the bottom row.}
\label{fig:solutions_B_dynamical_k}
\end{figure}

Finally, we show the lumps with higher topological charges $k=1,2,3,4,5,6$ for $\beta = 1$ in Fig.~\ref{fig:solutions_B_dynamical_k}.
The profiles of $\Theta$ are qualitatively the same as those for the constant magnetic field shown in Fig.~\ref{fig:solutions_B_const_k}. The gauge field $a$ and magnetic flux density $f_{12}$ show plateaus inside the lump cores. The lumps are no longer empty and are filled by the magnetic field. These behaviors in $f_{12}$ and ${\cal E}$ are similar to those of well-known Abrikosov-Nielsen-Olsen vortices.

\begin{figure}[t]
\begin{center}
\includegraphics[height=6.5cm]{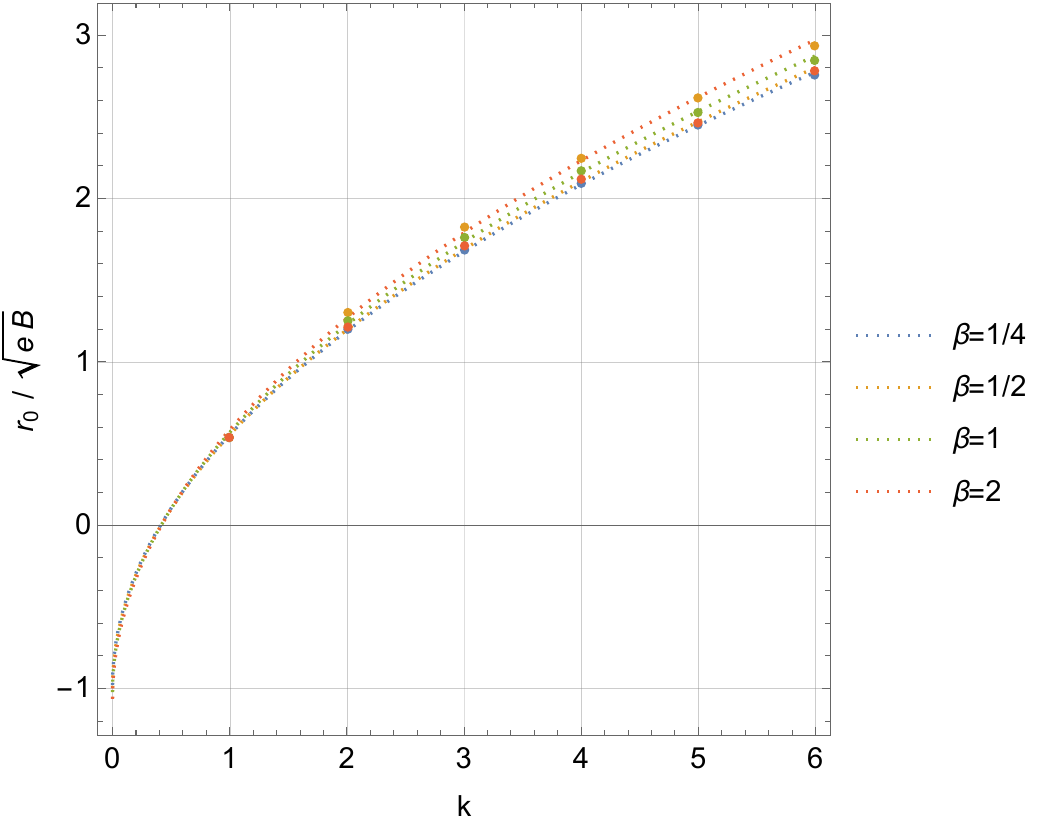}
\caption{The relation between the lump size $r_0$ and the winding number $k$ for $\beta = \{1/4,\,1/2,\,1,\,2\}$.}
\label{fig:dyn_gauge_r0_vs_k}
\end{center}
\end{figure}

We also numerically solved EOMs for $\beta = 1/4,\,1/2$ and $\beta = 2$, and the results are qualitatively the same as those for $\beta=1$.
As an example to see the similarity, we show the relation between the lump width $r_0$ and the winding number $k$ for $\beta = \{1/4,\,1/2,\,1,\,2\}$ in Fig.~\ref{fig:dyn_gauge_r0_vs_k}. We find it is well approximated by $\dfrac{r_0(k)}{\sqrt{eB}} = \sqrt{ak}+b$ as before. Namely, the lumps with magnetic fluxes behave as droplets.

\subsection{
Gauged anti-lumps in the down vacuum with $eB > 0$}
\label{sec:dyn_gauge_antilump}

Let us construct anti-lumps with the negative winding number $k<0$ which do not exist in the case of nondynamical gauge field.
For comparison, we take the same boundary condition as that taken in Sec.~\ref{sec:numerical_sol_dyn}. Namely, we consider the anti-lumps in the negative vacuum and the background magnetic field is fixed by $eB>0$. 

We first remind that, in the treatment that the $U(1)$ gauge field is non-dynamical in Sec.~\ref{sec:nondynamical}, the Derrick's scaling condition (\ref{eq:Derrick}) strictly prohibits the anti-lumps existing in the down vacuum with $eB>0$. On the other hand, in the treatment that the $U(1)$ gauge field is dynamical, the scaling condition is relaxed as Eq.~(\ref{eq:Derrick2}) due to the additional contribution by $E_4$. Therefore, there would exist a room for the anti-lumps to exist in the model with dynamical gauge fields. 

To verify this, we numerically solve EOMs in Eqs.~(\ref{eq:eom_full_1}) and (\ref{eq:eom_full_2}) with $k<0$ and the boundary condition $\Theta(0) = 0$ and $\Theta(\infty) = \pi$. Indeed, we numerically found non-singular anti-lump solutions.

\begin{figure}[ht]
\centering
\begin{minipage}[t]{0.49\linewidth}
\centering
\includegraphics[keepaspectratio,scale=0.4]{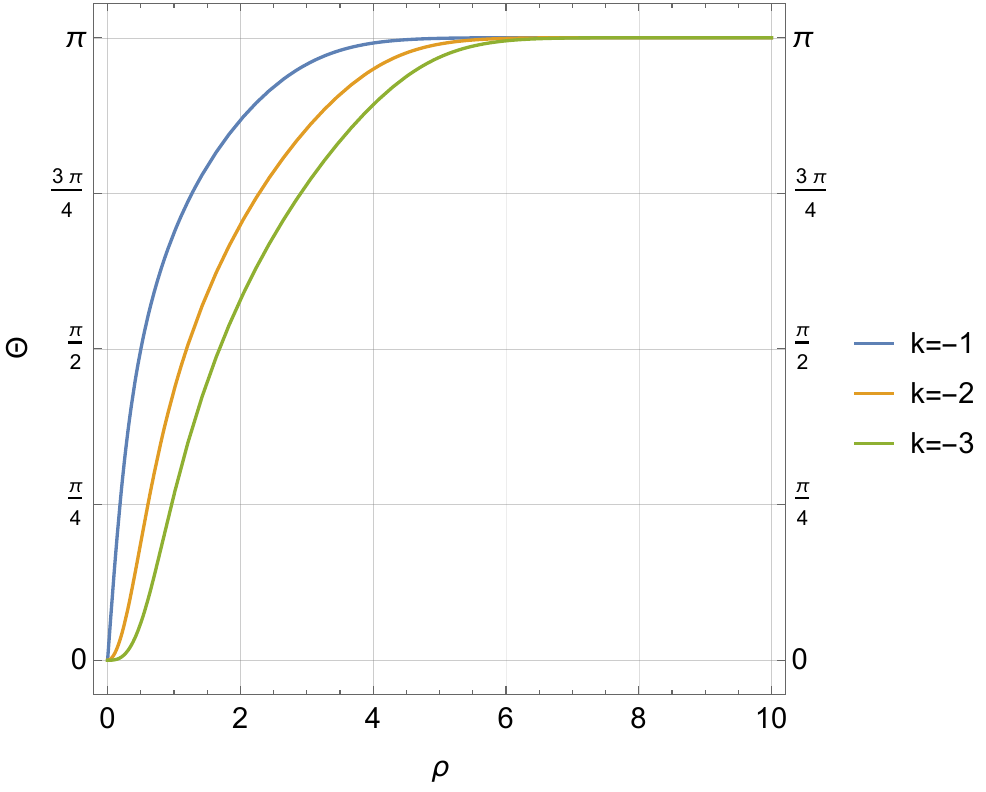}
\end{minipage}
\begin{minipage}[t]{0.49\linewidth}
\centering
\includegraphics[keepaspectratio,scale=0.4]{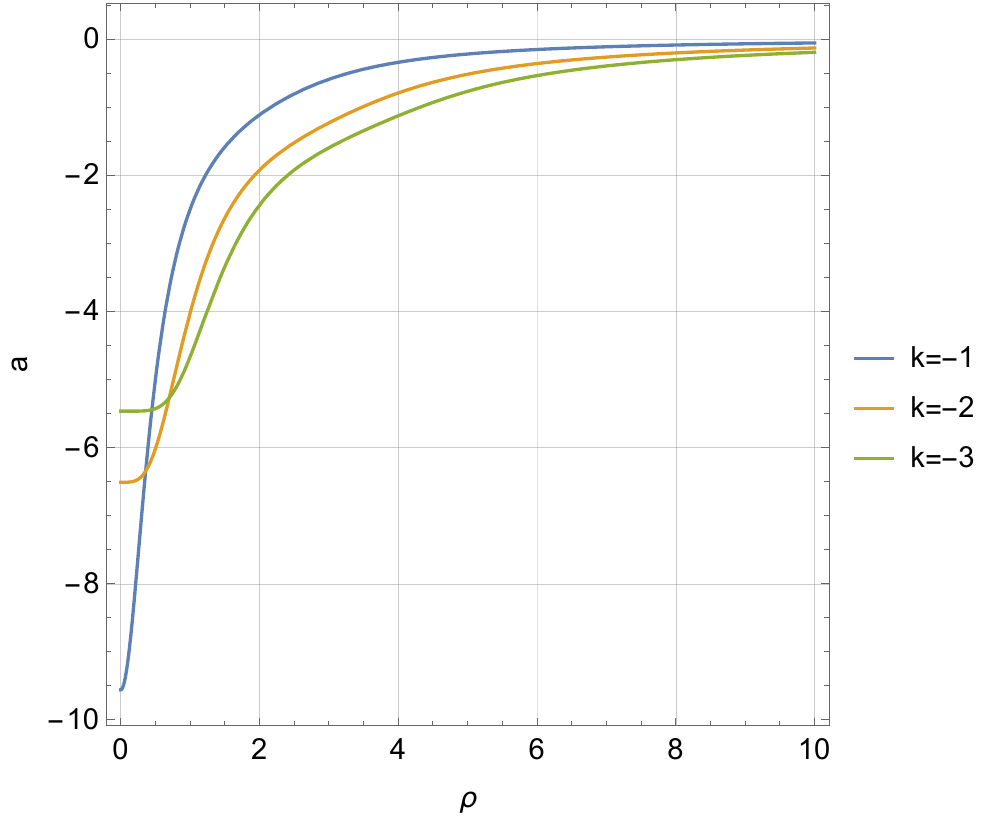}
\end{minipage}
\begin{minipage}[t]{0.49\linewidth}
\centering
\includegraphics[keepaspectratio,scale=0.4]{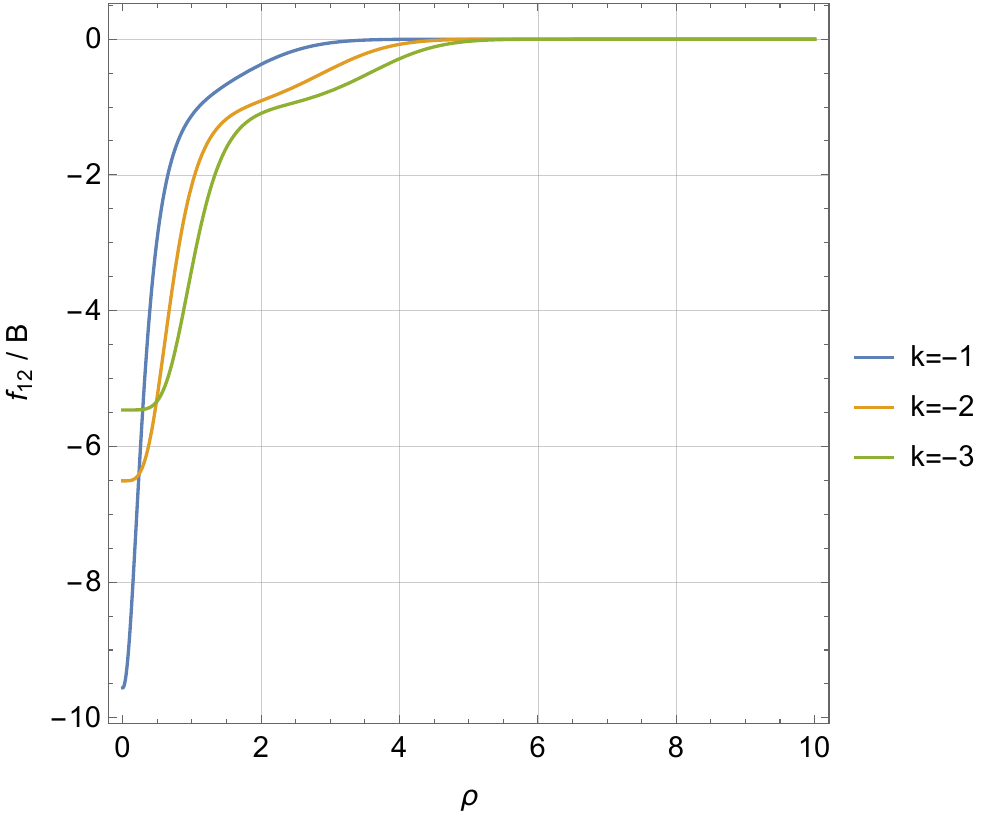}
\end{minipage}
\begin{minipage}[t]{0.49\linewidth}
\centering
\includegraphics[keepaspectratio,scale=0.4]{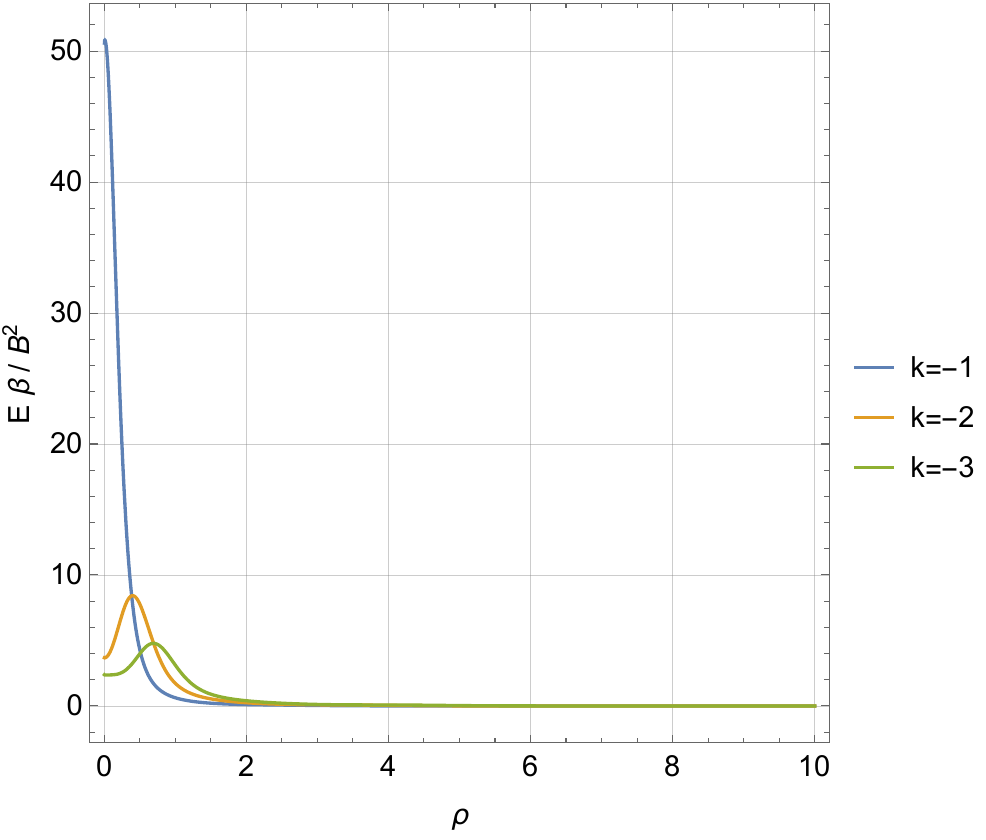}
\end{minipage}
\caption{The axially symmetric lumps for $k=-1,-2,-3$ for $\beta=1/4$ in the down vacuum. $\Theta$ and $a$ are shown in the top row, and $f_{12}$ and ${\cal E}$ are shown in the bottom row.}
\label{fig:solutions_B_dynamical_nk}
\end{figure}
As an example, we show the numerical solutions with $k=-1,\,-2,\,-3$ for $\beta = 1/4$ in Fig.~\ref{fig:solutions_B_dynamical_nk}.
The anti-lumps trap the negative magnetic field in the cores which is opposite to the background one $B$($>0$). Unlike the case with $k>0$, the back reaction $f_{12}$ is everywhere negative.

\begin{figure}[ht]
\begin{center}
\includegraphics[height=6.5cm]{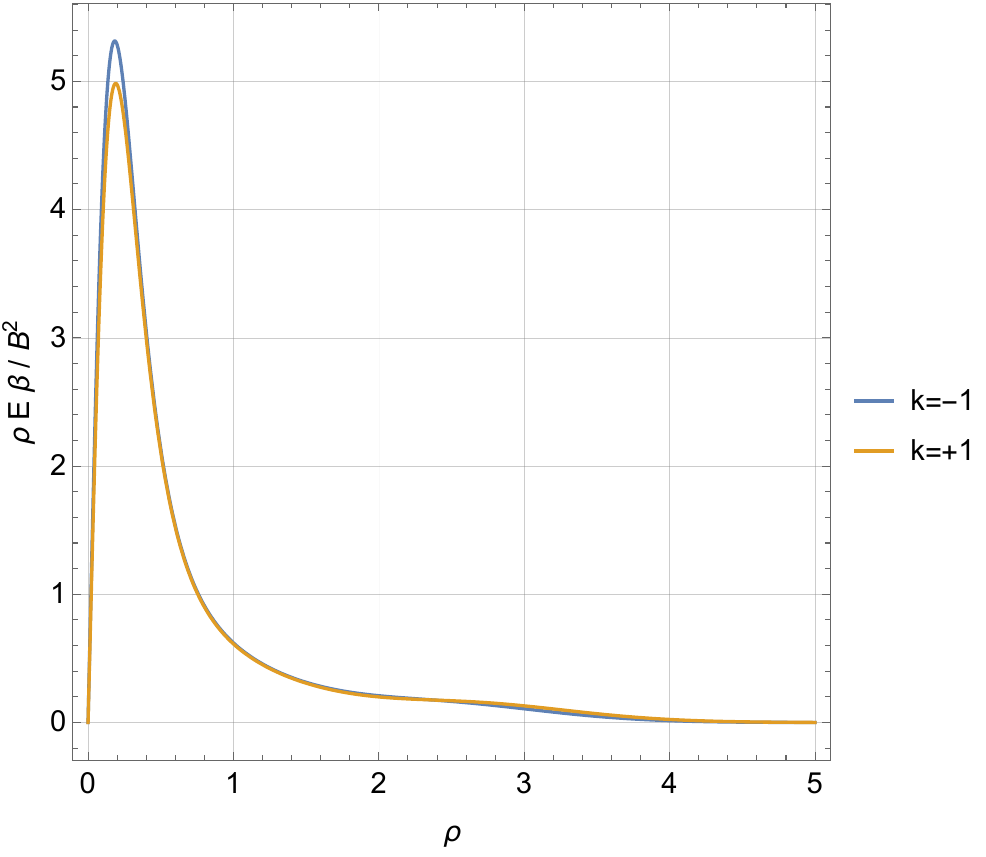}
\caption{The comparison of energy densities of $k=-1$ and $k=+1$ for $\beta = 1/4$.}
\label{fig:dyn_gauge_comparison_pk_vs_nk}
\end{center}
\end{figure}

The asymmetry between the lumps and anti-lumps in the model with the dynamical gauge field is more modest than that in the model without the dynamical field. However, the asymmetry does not disappear and remains as obviously seen in the EOMs in Eqs.~(\ref{eq:eom_full_1}) and (\ref{eq:eom_full_2}). This can be also verified by our numerical solutions. Fig.~\ref{fig:dyn_gauge_comparison_pk_vs_nk} shows the energy densities (more explicitly, we show ${\cal E}\rho \beta/B^2$) of the $k=-1$ and $k=+1$ lumps under the same down vacuum with $\beta = 1/4$. We find that $k=-1$ lump has higher energy around its core and slightly lower at the outer bump. We numerically evaluate the mass of $k=-1$, and found $\dfrac{M}{2\pi v^2} = 2.97$ which is higher than $\dfrac{M}{2\pi v^2} = 2.90$ for $k=+1$, as shown in Table \ref{table}.

In conclusion, both the lumps and anti-lumps exist in the down vacuum with $eB>0$, but the former is energetically favored than the latter. Intuitively, this reflects the fact that putting the soliton with the magnetic field opposite to the vacuum costs an energy than the soliton with the same magnetic field as the vacuum. The mass difference becomes larger as $\beta$ is increased. In the limit $\beta \to \infty$ where the gauge field is non-dynamical, the anti-lumps tend to be singular and to be excluded from the system.

The same can be said for the up vacuum with $eB >0$ or $eB<0$ by appropriately exchanging the lump and anti-lump.

\section{Summary and discussion} \label{sec:summary}

In this paper,
we have worked out stable lumps in
a gauged $O(3)$ model without any potential term 
coupled with a (non)dynamical $U(1)$ gauge field.
We have found that 
the gauged-lumps without a potential term can be made stable 
by putting them in a uniform magnetic field, 
irrespective of whether the gauge field is dynamical or not. 
In the case of the non-dynamical gauge field,  
we have found that only either of lumps or anti-lumps stably exists depending on the sign of the background magnetic field, and the other is unstable to shrink to be singular. 
The case of nondynamical gauge field resemble magnetic Skyrmions in chiral magnets.
We have also constructed coaxial lumps with higher winding number $k$,  
whose size and mass are proportional to $\sqrt{k}$ up to a constant shift, implying that they behave as droplets. 
On the other hand, in the case of the dynamical gauge field, 
we have found that both the lumps and anti-lumps 
stably exist  with different masses; 
the lighter (heavier) one corresponds to 
the (un)stable one in the case of the nondynamical gauge field.
The lumps behave as superconducting rings,   
trapping magnetic fields in their interiors. 
The total magnetic fluxes are reduced from 
the background magnetic field because of
superconductivity.

In this paper, we have discussed only axially symmetric configurations for multiple lump states. 
The next step should be to investigate 
separated configurations and study interaction 
among them. 
Since the profile of a single lump exponentially decays, 
the interaction would be exponentially suppressed 
like vortices in superconductors.
In particular, we expect a repulsive force between two lumps and the existence of a lattice 
as the case of magnetic Skyrmions \cite{Rossler:2006,doi:10.1126/science.1166767,doi:10.1038/nature09124,Han:2010by,Ross:2020hsw}\footnote{
However, the interaction between magnetic Skyrmions was polynomial like \cite{Ross:2020hsw}.
}.
This is quite important for 
an application to domain-wall Skyrmions in 
QCD under strong magnetic field.

One natural generalization of the current work is the ${\mathbb C}P^{N-1}$ model. From a theoretical point of view, 
a naive question is if one $U(1)$ gauging is enough or 
$U(1)^{N-1}$ gauging is necessary  for 
the stability of ${\mathbb C}P^{N-1}$  lumps. 
From a physical point of view,  
such a $U(1)$ gauged ${\mathbb C}P^2$ 
model appears 
on a chiral soliton with three flavors (up, down and strange quarks) 
in which case ${\mathbb C}P^2$ lumps are 
$SU(3)$ Skyrmions from the bulk point of view.
The $U(1)$ gauged ${\mathbb C}P^2$ 
model also appears \cite{Vinci:2012mc,Hirono:2012ki,Chatterjee:2015lbf} 
on the worldsheet of 
a non-Abelian vortex in dense QCD \cite{Balachandran:2005ev,Nakano:2007dr,
Eto:2009kg,Eto:2009bh,Eto:2009tr,Eto:2013hoa}
in which case lumps are sigma model instantons 
viewed as Yang-Mills instantons from the bulk 
\cite{Eto:2004rz}.

Our model would have an impact on production of topological solitons.
During a phase transition accompanied with a spontaneous symmetry breaking, 
topological solitons are in general created by the
Kibble-Zurek-mechanism.
Usually the numbers of solitons and anti-solitons 
created in this way are the same.
If we consider such a phase transition 
in our model, 
the background magnetic field works as a bias leading to an asymmetry between the lumps or anti-lumps. 
Hence, this suggests a new impact on solitogenesis and 
might be useful for baryogenesis in early Universe.

\begin{acknowledgments}
YA is grateful for the hospitality at the Department of Physics of Yamagata University, where this work was initiated.
This work is supported in part by 
 JSPS KAKENHI [Grants  No.~JP23KJ1881 (YA), No. JP22H01221 (ME and MN)] and the WPI program ``Sustainability with Knotted Chiral Meta Matter (SKCM$^2$)'' at Hiroshima University (ME and MN).
 
\end{acknowledgments}

\clearpage

\appendix

\section{A BPS lump for $B=0$}
\label{apndx:BPS_lump}

Here let us quickly recall the analytic solution for $B=0$. The EOM reduces to one for a usual $\mathbb{C}P^1$ lump, and its analytic solution is available as
\begin{eqnarray}
    \frac{n_3(r)}{v} = \cos\Theta(r) =  \frac{a^2 - r^2}{a^2 + r^2}\,.
\end{eqnarray}
Here, $a$ is so called size modulus. As usual, we define the size $r_0$ of lump by $n_3(r_0) = 0$. Clearly, we have $r_0 = a$.
This solution satisfies the boundary condition $\cos\Theta(0)=0$ and $\cos\Theta(\infty) = \pi$ ($n_3(0)=v$  and $n_3(\infty)=-v$).
The asymptotic behavior at $r \to \infty$ is power law as $\frac{n_3}{v} = -1 + \frac{2a^2}{r^2} + \cdots$.

Let $C$ be a closed curve on which $n_3 = 0$, 
and $D$ be the interior of $C$. The $U(1)$ is 
spontaneously and maximally broken around $|n_1+ i n_2|=v$. 
Thus, the closed curve 
$C$ is a superconducting ring, 
and there is a persistent current along it.
Let us write $n_1+ i n_2 = v e^{i\psi}$ on $C$. 
The configuration of the gauge field 
along $C$ is determined by minimizing 
the gradient energy 
$|D_\alpha (n_1+ i n_2)|^2 = 0$, 
yielding $\partial_\alpha \psi = e A_\alpha$.
Then, we have a flux (and area) 
quantization on $D$:
\begin{eqnarray}
B S_D= \int_D  d^2x\, B = \oint_C  d x^i A_i
= \frac{1}{e} \oint_C  d x^i \partial_i \psi
= \frac{2\pi}{e} \;\; \label{quantize}
\end{eqnarray}
with the area $S_D$ of $D$.
This gives a constraint among the lump moduli.
For a single BPS lump with $n_3 = v (a^2-r^2)/(a^2+r^2)$.
Thus, the size of $D$ bounded by $n_3=0$ 
is $r=a$, 
and the flux quantization implies
a quantization of the size $a=\sqrt{2/eB}$. 

\bibliographystyle{jhep}
\bibliography{reference}

\end{document}